# The Fundamental Plane for cluster E and S0 galaxies*

Inger Jørgensen[1,2,4,†], Marijn Franx[3,4], and Per Kjærgaard[1]
[1] *Copenhagen University Observatory, Øster Voldgade 3, DK-1350 Copenhagen, Denmark*
[2] *McDonald Observatory, The University of Texas at Austin, RLM 15.308, Austin, TX 78712, USA (Postal address for IJ)*
[3] *Kapteyn Institute, P.O.Box 800, 9700 AV Groningen, The Netherlands (Postal address for MF)*
[4] *Center for Astrophysics, 60 Garden Street, Cambridge, MA 02138, USA*
e-mail: inger@roeskva.as.utexas.edu, franx@cosmos.astro.rug.nl, per@astro.ku.dk

**ABSTRACT**
We have analyzed the shape of the Fundamental Plane (FP) for a sample of 226 E and S0 galaxies in ten clusters of galaxies. We find that the distribution of galaxies is well approximated by a plane of the form

$$\log r_e = 1.24 \log \sigma - 0.82 \log <I>_e + \gamma$$

for photometry obtained in Gunn r. This result is in good agreement with previous determinations. The FP has a scatter of 0.084 in $\log r_e$. For galaxies with velocity dispersion larger than $100 \text{km s}^{-1}$ the scatter is 0.073. If the FP is used for distance determinations this scatter is equivalent to 17% uncertainties on distances to single galaxies.
We find that the slope of the FP is not significantly different from cluster to cluster. Selection effects and measurement errors can introduce biases in the derived slope. The residuals of the FP correlate weakly with the velocity dispersion and the surface brightness. Some of the coefficients used in the literature give rather strong correlations between the residuals and absolute magnitudes. This implies that galaxies need to be selected in a homogeneous way to avoid biases of derived distances on the level of 5–10% or smaller.
The FP has significant intrinsic scatter. No other structural parameters like ellipticity or isophotal shape can reduce the scatter significantly. This is in contradiction to simple models, which predict that the presence of disks in E and S0 galaxies can introduce scatter in the FP. It remains unknown what the source of scatter is. It is therefore unknown whether this source produces systematic errors in distance determinations.
The $Mg_2$-$\sigma$ relation for the cluster galaxies differs slightly from cluster to cluster. Galaxies in clusters with lower velocity dispersions have systematically lower $Mg_2$. The effect can be caused by both age and metallicity variations. With the current stellar population models, it is in best agreement with our results regarding the FP if the offsets are mainly caused by differences in metallicity.
Most of the distances that we derive from the FP imply small peculiar motions (< $1000 \text{km s}^{-1}$). The zero point of the FP must therefore be quite stable. Only for one cluster, located 28° from the direction towards the "Great Attractor", we find a peculiar motion of $1300 \text{km s}^{-1}$. This motion is reduced to $890 \text{km s}^{-1}$ if we use the FP corrected for the offset of the $Mg_2$-$\sigma$ relation. This confirms earlier suggestions that the residuals from the $Mg_2$-$\sigma$ relation can be used to flag galaxies with deviant populations, and possibly to correct the distance determinations for the deviations.

---

* Based in part on observations obtained at the European Southern Observatory, La Silla, Chile
† Hubble Fellow

## 1 INTRODUCTION

The question of how to characterize E galaxies and how many parameters are necessary for the characterization is important for our understanding of the galaxies as a class.

Terlevich et al. (1981) and de Vaucouleurs & Olson (1982) were the first to point out that E galaxies form at least a two-parameter class. Until then the luminosity was thought to be the determining parameter for the galaxy properties.

With the work on E galaxies by Djorgovski & Davies (1987) and by Dressler et al. (1987b) it became clear that the effective radius (or the total luminosity), the central velocity dispersion, and the mean surface brightness are related in a relation which is linear in the logarithmic space. This is the Fundamental Plane (FP): $\log r_e = \alpha \log \sigma + \beta \log <I>_e + \gamma$. The relation has very low scatter, 15–20% in the radius, and is, therefore, compared to the Faber-Jackson relation (Faber & Jackson 1976) a major improvement for distance determinations.

The FP is thought to originate from the formation and evolution processes of the galaxies. However, the exact physical origin remains unclear. If the luminosity profiles as well as the dynamical structure of the galaxies are similar then the virial theorem implies that the mass-to-light (M/L) ratio is a function of $r_e$, $<I>_e$, and $\sigma$. If this function is unique and a power law, then a relation like the FP is to be expected (see also Djorgovski, de Carvalho, & Han 1989; Faber et al. 1987). Which physical processes that cause the M/L ratio to be a unique function of $r_e$, $<I>_e$, and $\sigma$ is presently not known. The existence and appearance of the FP imply that the M/L ratio depends on the mass of the galaxy (e.g. Bender, Burstein, & Faber 1992).

The $D_n$-$\sigma$ relation introduced by Dressler et al. (1987b) is closely related to the FP. For normal E galaxies Dressler et al. found $\log D_n/r_e = -0.8 \log <I>_e +$ constant, thus that the $D_n$-$\sigma$ relation gives a nearly edge-on view of the FP.

The interest in relations like the FP and the $D_n$-$\sigma$ relation between global parameters for galaxies is twofold. The relations provide information on the properties of the galaxies as a class, and the relations may be used for distance determination. The application of the relations as distance determinators is based on the assumption that they at least within some accuracy are universally valid. Dressler and collaborators used the $D_n$-$\sigma$ relation as distance determinator and found peculiar velocities relative to the Hubble-flow of 500-1000km s$^{-1}$ and even larger in the Hydra-Centaurus supercluster (Dressler et al. 1987a; Lynden-Bell et al. 1988; Faber et al. 1989). They proposed the existence of a huge mass-concentration, the so-called "Great Attractor", as the source of the flow (Lynden-Bell et al.; Faber & Burstein 1988). The interpretation of the current data on peculiar velocities is, however, still uncertain. Mathewson, Ford, & Buchhorn (1992) did not detect the expected backside infall towards the "Great Attractor", and recently Courteau et al. (1993) found evidence for a bulk flow of all galaxies within 6000km/s. Mould et al. (1993) investigated the flow on basis of distances derived from the Tully-Fisher relation (Tully & Fisher 1977), and could not distinguish between bulk flow models and models with a "Great Attractor".

It has also been a topic of discussion if the FP and the $D_n$-$\sigma$ relation have universal validity. In order to interpret the derived peculiar velocities correctly we need to understand the methods, and especially the possible systematic effects in the methods.

One can make the following hypotheses about the E and S0 galaxies in clusters, and a generalized form of the FP which may include a correction term for variations in the stellar populations:

(a) All E and S0 galaxies in clusters are similar in all properties in the sense that they are drawn from the same probability function. The probability function is a function of the M/L ratio, $r_e$, $<I>_e$, $\sigma$, $Mg_2$, and probably other characteristic parameters for the galaxies. A consequence of this hypothesis is that if a (generalized) FP exists, then it is universally valid. Further, the intrinsic scatter for the FP and the distribution within the FP will be the same for all clusters.

(b) The E and S0 galaxies are not drawn from the same probability function, but they all follow the (generalized) FP and the distribution perpendicular to the FP is the same for all clusters. In this case the scatter for the FP is the same for all clusters. The distribution within the FP is, however, dependent on other properties, e.g. properties of the environment.

(c) The E and S0 galaxies are not drawn from the same probability function, and the distribution perpendicular to the FP is not the same for all clusters. The FP may be valid within certain limits and may be used for distance determinations. The scatter is, however, not the same for all clusters, and asymmetric residuals for different clusters may cause systematic errors in the distance determinations.

It is possible that well-defined subclasses of E and S0 galaxies, e.g. with upper and lower limits in luminosity, satisfy the strongest hypothesis (a), while this may not the case for the whole population.

Several authors have discussed topics related to these hypotheses, and similar hypotheses for field galaxies as well. Djorgovski et al. (1989) found the slope of the $D_n$-$\sigma$ relation to depend on the cluster richness. A similar discussion by Burstein, Faber, & Dressler (1990) shows that this is not the case. Recently de Carvalho & Djorgovski (1992) have found cluster and field E galaxies to be different, and that this difference shows up in the FP. The field galaxies have lower $Mg_2$ line index and higher surface brightness. A similar effect was found by Guzmán et al. (1992) for the E galaxies in the outer parts of the Coma cluster. Also Gregg (1992) found the stellar populations in the galaxies to be correlated with the residuals from the $D_n$-$\sigma$ relation. Bender et al. (1992) discussed the importance of the FP for galaxy evolution. The authors found differences in loci within the FP to be related to differences in velocity structure and evolutionary history.

Because both the interpretation of the data on distances and peculiar velocities, and the methods used for the determination of these parameters are subject to disputes, we have undertaken an observing program in order to provide new independent data of E and S0 galaxies.

The first CCD photometry from this program covered galaxies in the Coma cluster (Jørgensen, Franx, & Kjærgaard 1992). Together with spectroscopic data from the literature this photometry was used in to show that the FP for E galaxies in the Coma cluster has a lower scatter than the $D_n$-$\sigma$ relation, and that the $D_n$-$\sigma$ relation in some cases may give systematic errors in the distance determinations (Jørgensen, Franx, & Kjærgaard 1993, hereafter Paper I). We, therefore, prefer the FP as the distance determinator.

In this paper we test the hypotheses about E and S0 galaxies in clusters and the FP for these galaxies, which were

The Fundamental Plane    3TABLE 1
Information on the clusters

| Cluster | $cz_{hel}$ [km/s] | $cz_{CMB}$ [km/s] | $cz_{LSR}$ [km/s] | B-M type | Richness | $\sigma_{cluster}$ [km/s] | $N_{gal}$ | Ref | $T_{gas}$ [keV] | Ref |
|---|---|---|---|---|---|---|---|---|---|---|
| Coma $\equiv$ Abell 1656 | 6922 | 7200 | 6931 | II | 106 | $1010^{+51}_{-44}$ | 234 | Z90 | 8.3$\pm$0.6 | D93 |
| Abell 194 | 5340 | 5038 | 5423 | II | 37 | $440^{+40}_{-30}$ | 74 | SR91 | 2$\pm$1 | JF84 |
| Abell 539 | 8741 | 8734 | 8664 | III | 50 | $701^{+61}_{-49}$ | 86 | SR91 | 3.0$\pm$0.5 | D93 |
| Abell 3381 $\equiv$ DC0608-33 | 11381 | 11472 | 11141 | I | 69 | $372^{+58}_{-39}$ | 32 | DS88 | | |
| Abell 3574 $\equiv$ Klemola 27 | 4665 | 4931 | 4491 | I | 31 | $479^{+70}_{-49}$ | 35 | W91 | | |
| Abell S639 | 6269 | 6545 | 5979 | I-II | 14 | $456^{+83}_{-54}$ | 24 | Paper IV | | |
| Abell S753 $\equiv$ NGC5419 group | 4174 | 4421 | 4000 | I | 18 | $544^{+79}_{-55}$ | 36 | W91 | | |
| DC2345-28 $\equiv$ Klemola 44 $\equiv$ A4038 | 8775 | 8482 | 8808 | III | 117 | $852^{+110}_{-79}$ | 43 | LC88 | 3.3$\pm$0.8 | D93 |
| Doradus $\equiv$ Grm13 | 1127 | 1104 | 914 | | 14 | $244^{+30}_{-22}$ | 46 | M89 | | |
| HydraI $\equiv$ Abell 1060 | 3718 | 4050 | 3450 | III | 39 | $608^{+58}_{-39}$ | 105 | Z90 | 3.9$\pm$0.2 | D93 |
| Grm15 | 4675 | 4575 | 4602 | | 9 | $487^{+149}_{-77}$ | 11 | M89 | | |

Note.— B-M – Bautz-Morgan type as given by Abell et al. (1989). Richness – The number of galaxies with magnitudes in the interval $m_3$ to $m_3 + 2$ as listed by Abell et al. Values for Doradus and Grm15 are calculated from the listings by Maia et al. (1989). $\sigma_{cluster}$ – the line of sight velocity dispersion of the cluster in the rest frame of the cluster. References for radial velocities and $\sigma_{cluster}$: DS88: Dressler & Shectman (1988), LC88: Lucey & Carter (1988), M89: Maia et al., Z90: Zabludoff et al. (1990), SR91: Strubble & Rood (1991), W91: Willmer et al. (1991). $\sigma_{cluster}$ has been calculated from the velocity data from DS88, W91, LC88, and Paper IV. Uncertainties are derived following Danese et al. (1980). $N_{gal}$ – the number of galaxies used for the determination of $\sigma_{cluster}$ and $cz_{hel}$. $cz_{hel}$ for the Coma cluster has been adapted from Faber et al. (1989). References for intracluster gas temperatures derived from X-ray observations: JF84: Jones & Forman (1984), D93: David et al. (1993).

TABLE 2
Number of galaxies with available data

| Cluster | Photometry | | | Spectroscopy | | | Phot.+Spec. | | | $R^a$ [Mpc $h^{-1}$] | Passbands for the photometry | Source of the spectroscopy |
|---|---|---|---|---|---|---|---|---|---|---|---|---|
| | $N_E$ | $N_{S0}$ | $N_{S+U}$ | $N_E$ | $N_{S0}$ | $N_{S+U}$ | $N_E$ | $N_{S0}$ | $N_{S+U}$ | | | |
| Coma | 53 | 93 | 25 | 44 | 37 | 0 | 44 | 37 | 0 | 1.03 | r, B | D87, Dav87, L91, G92 |
| A194 | 8 | 16 | 4 | 8 | 14 | 2 | 8 | 14 | 2 | 1.04 | r, g, B, U | Paper IV, LC88 |
| A539 | 8 | 18 | 1 | 6 | 23 | 2 | 6 | 18 | 1 | 0.48 | r, g | Paper IV |
| A3381 | 8 | 10 | 2 | 6 | 19 | 6 | 6 | 8 | 0 | 1.45 | r, g | Paper IV |
| A3574 | 4 | 15 | 3 | 1 | 6 | 0 | 1 | 6 | 0 | 0.86 | r, B | Paper IV |
| S639 | 6 | 4 | 3 | 6 | 4 | 14 | 6 | 4 | 2 | 0.21 | r, g | Paper IV |
| S753 | 5 | 14 | 2 | 5 | 9 | 1 | 5 | 9 | 1 | 0.78 | r, B | Paper IV |
| DC2345-28 | 11 | 21 | 3 | 11 | 18 | 0 | 11 | 18 | 0 | 0.33 | r, g, B, U | Dav87, LC88 |
| Doradus | 3 | 6 | 1 | 3 | 5 | 1 | 3 | 5 | 1 | 2.55 | r, g | Paper IV |
| HydraI | 8 | 16 | 0 | 13 | 29 | 0 | 7 | 12 | 0 | 1.07 | r, g | Paper IV |
| Grm15 | 3 | 2 | 3 | 2 | 2 | 3 | 2 | 2 | 3 | 1.09 | r, g | Paper IV |
| Total | 117 | 215 | 47 | 105 | 166 | 29 | 99 | 133 | 10 | ... | ... | ... |

Note.— The number of galaxies in each class takes into account our reclassifications on basis of CCD photometry (Paper III). The number of spirals and unclassfied galaxies is given as $N_{S+U}$. Non-members are not included. $^a$ Interval in cluster center distance spanned by the observed galaxies. One galaxy in A194 at R=2.36Mpc $h^{-1}$ has been observed. Ref. – D87: Dressler (1987), Dav87: Davies et al. (1987), LC88: Lucey & Carter (1988), L91: Lucey et al. (1991), G92: Guzmán et al. (1992).

outlined above. Specific attention is paid to test for dependences of the cluster properties and to investigate possible sources of the intrinsic scatter. The analysis is done on basis of photometry and spectroscopy for galaxies in eleven clusters. The relative distances and peculiar velocities for the clusters seen in connection with current models for the large-scale flow will be discussed in a later paper.

The sample selection and the data are shortly described in Sect. 2. The FP in Gunn r is derived in Sect. 3 and the distribution within the FP and perpendicular to the FP are investigated. This section also includes a discussion of the correlations between the M/L ratio and other observables. Environmental effects are studied in Sect. 4. The dependence on galaxy type and effects of projection and differences in relative disk luminosities are discussed in Sect. 5. The importance of the stellar populations is investigated in Sect. 6. In Sect. 7 we shortly discuss the FP as a distance determinator. The conclusions are summarized in Sect. 8.

## 2   SAMPLE SELECTION AND DATA

The clusters for this project were selected to span a wide range in richness and regularity. Information on the clusters is summarized in Table 1. The observed galaxies were drawn from lists of known E and S0 galaxies in these clusters. The limiting magnitude varied from cluster to cluster. The five brightest galaxies of each cluster were observed, and the other galaxies were chosen at random from the lists.

We obtained spectroscopy and/or photometry for 371 E and S0 galaxies in eleven clusters (Jørgensen & Franx 1994, hereafter Paper II; Jørgensen, Franx, & Kjærgaard 1995ab, hereafter Paper III and IV). Photometry is available for 80% of the galaxies in the original sample outside of Coma, and spectroscopy for 70%. The spectroscopy includes both our observations as well as velocity dispersions and $Mg_2$ line indices from literature. For the Coma cluster a complete magnitude limited sample of 146 E and S0 galaxies has photometry (Paper II). Spectroscopy has been published for 55% of these. The selection criteria are described in more detail in Paper III and IV. In total 232 E and S0 galaxies in the eleven clusters have observed photometric and spectroscopic parameters, $r_e$, $<I>_e$ and $\sigma$. Here $r_e$ is the effective radius, $<I>_e$ the mean surface brightness within this radius, and $\sigma$ the (central) velocity dispersion of the galaxy.

The velocity dispersions and the $Mg_2$ indices have been corrected for the size of the aperture. Fully corrected values for both our data and the literature data are given in Paper IV. In this paper we use velocity dispersion and $Mg_2$ corrected to a circular aperture with diameter $1.19h^{-1}$ kpc. This is equivalent to $3\rlap.{''}4$ at the distance of the Coma cluster. For comparison we sometimes use spectroscopic parameters corrected to a circular aperture with diameter $r_e/4$. Table 2 lists the number of galaxies with available data in each cluster and the sources of data.

The measurement errors of the photometric and spectroscopic observations for our data are given in Paper III and IV. The typical measurement errors are: $\log r_e$: $\pm 0.045$; $\log <I>_e$: $\pm 0.064$; $\log L$: $\pm 0.036$; $\log \sigma$: $\pm 0.036$; and $Mg_2$: $\pm 0.013$. The mean surface brightness $<I>_e$ is measured in $L_\odot/pc^2$. $<I>_e$ is calculated from the mean surface brightness in mag arcsec$^{-2}$ as $\log <I>_e = -0.4(<\mu>_e - \text{constant})$.

The constants for the passbands Gunn r, g, Johnson B, and U are 26.4, 26.45, 27.0 and 27.15, respectively. The errors in $\log r_e$ and $\log <I>_e$ are highly correlated, and the uncertainty on the combination $\log r_e + 0.82 \log <I>_e$ which enters the FP is $\pm 0.020$. The typical measurement errors on the literature data are: $\log \sigma$: $0.025 - 0.036$; and $Mg_2$: $\pm 0.010$. In Paper IV we found that measurements of velocity dispersions below $100 \text{km s}^{-1}$ may be affected by systematic errors. However, since the determination of the FP can be heavily biased by imposing a cut in the velocity dispersion at $100 \text{km s}^{-1}$ (cf. Sect. 3.1) we have not excluded the low velocity dispersion galaxies from the analysis.

## 3   THE FUNDAMENTAL PLANE IN GUNN R

The FP in Gunn r has been determined from the ten clusters with seven or more observed early-type galaxies. The cluster Grm15 was excluded, because it turned out to contain only four observed early-type galaxies. The galaxies Coma-D120 and Coma-D121 were excluded from the determination, as the angular distance between these galaxies is so small that the photometric parameters are highly uncertain. 226 E and S0 galaxies were included in the determination of the FP.

We fitted a plane to the distribution of galaxies in $(\log r_e, \log <I>_e, \log \sigma)$. This was done as an "orthogonal fit"; we seek the vector normal to the plane, $\vec{n} = (-1, \alpha, \beta)$, that minimizes the sum of the absolute residuals perpendicular to the plane. The residual perpendicular to the plane can be written as $\Delta = (\log r_e - \alpha \log \sigma - \beta \log <I>_e)/(1+\alpha^2+\beta^2)^{1/2}$. This fitting technique gives a normal to the FP defined by $\vec{n} = (-1, \alpha, \beta) = (-1, 1.24, -0.82)$. The equivalent equation for the FP is

$$\log r_e = \begin{array}{c} 1.24 \log \sigma \\ \pm 0.07 \end{array} \begin{array}{c} - 0.82 \log <I>_e \\ \pm 0.02 \end{array} + \gamma_{cl} \quad (1)$$

The uncertainties of the coefficients were derived by a bootstrap procedure. The uncertainties are discussed further below. The relative distances of the clusters were left free, and the median zero point, $\gamma_{cl}$, was used for each cluster. The median zero points are directly related to the relative distances of the clusters.

The fact that we minimize the sum of the absolute residuals instead of using the least squares method has the advantage that the procedure is relatively insensitive to a few outliers. Also, we use the median zero points, $\gamma_{cl}$, for the clusters instead of the mean zero points. The fitting procedure treats the parameters symmetrically. This is to preferred when there are measurement errors in all parameters, and our goal is to determine the physical relation between the parameters.

Figure 1a shows the face-on view of the FP given by $x = (2.21 \log r_e - 0.82 \log <I>_e + 1.24 \log \sigma)/2.66$, $y = (1.24 \log <I>_e + 0.82 \log \sigma)/1.49$. The x-axis is proportional to $\log r_e$. As previously pointed out by Guzmán, Lucey & Bower (1993) it is clear that the galaxies occupy only a small area in this plane. The distribution is limited by the selection effect in the absolute magnitude. The dashed line on the figure marks the absolute magnitude $-20\rlap.{^m}45$ in Gunn r, the magnitude limit for the Coma cluster sample. However, the upper boundary, $y \approx -0.54x + 4.2$ (dotted on the figure), is not created by selection effects. This boundary is the same as so-called the exclusion zone noted by Bender et al. (1992).

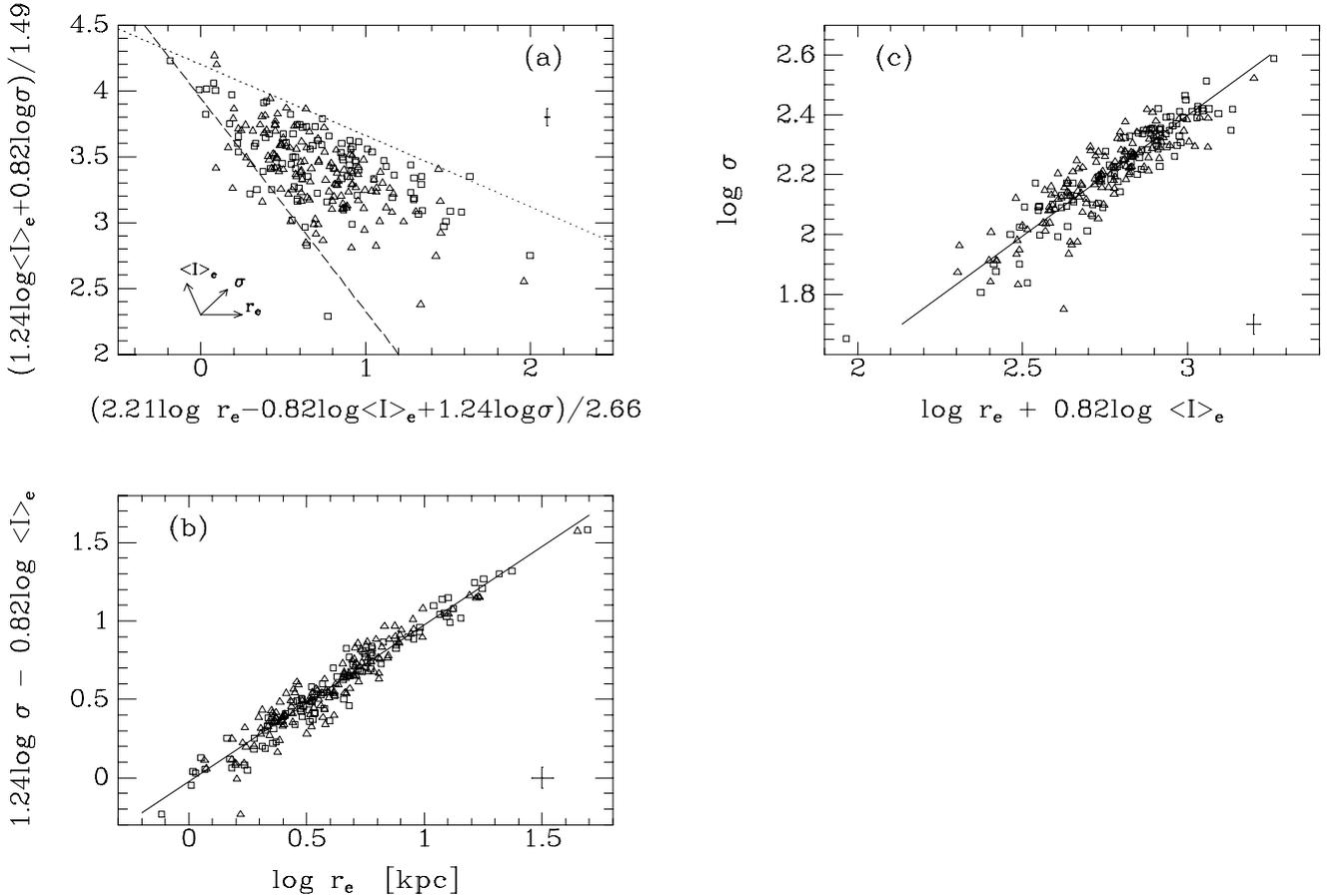

**Figure 1.** (a) The FP in Gunn r as derived in Eq. 1 shown face-on for all galaxies in the sample. The arrows in the lower left corner of the panel show in which directions the measured parameters increase. The dashed line shows the selection effect due to a limiting magnitude of $-20^m 45$ in Gunn r. This is the magnitude limit for the Coma cluster sample. The upper boundary (dotted line, $y \approx -0.54x + 4.2$) is not caused by selection effects. (b) The FP in Gunn r (Eq. 1) shown edge-on for all galaxies in the sample. This edge-on view is along one of the longest sides of the plane, the effective radius. (c) The FP in Gunn r (Eq. 1) shown edge-on along one of the shortest sides of the plane, the velocity dispersion. The effective radii have been derived in kiloparsec from the relative distances given in Table 4; $H_0 = 50$ km s$^{-1}$ Mpc$^{-1}$ was used. Boxes – E galaxies, triangles – S0 galaxies. Typical error bars are given on the panels.

Figure 1b shows the FP edge-on. The axis in the horizontal direction, $\log r_e$, is one of the longest of the distribution. This makes the figure "look good". Figure 1c shows the FP edge-on in a different direction, along the "shortest" axis of the plane. Thus, the scatter is more apparent. The FPs for the individual clusters are shown in Figure 2.

Figure 1b and c seem to indicate that a small curvature of the FP may be present. The largest galaxies may lie slightly below the plane. If we fit the coordinate along the FP, $x$, directly to the coordinate perpendicular to the FP, $z = (-\log r_e - 0.82\log <I>_e + 1.24\log \sigma)/1.79$, and allow for a second order term in $z$ the significance of the second order term is on the $2\sigma$ level. The scatter is only reduced marginally by inclusion of a second order term. If the coefficients for the FP are changed the curvature can become stronger. This is due to the fact that the galaxies lie in a triangular area within the FP. If the coefficients are changed this area gets tilted. The tilt introduces small second-order effects. From the current data, we conclude that there is no strong evidence for departures from a plane. A data set spanning a larger range in absolute magnitudes is needed to address this question in more detail.

We have also derived the FP using the velocity dispersion corrected to an aperture that encloses a fixed fraction of the luminosity. We have used a circular aperture with diameter $r_e/4$. The best fitting plane is $\log r_e = 1.30 \log \sigma - 0.87 \log <I>_e + \gamma_{cl}$. The aperture correction implies that $\log \sigma(1.19h^{-1}$ kpc$) = \log \sigma(r_e/4) + 0.04 \log r_e +$ constant (Paper IV). Thus, the FP derived with $\log \sigma(r_e/4)$ is in agreement with Eq. 1 and the aperture correction.

### 3.1 Comparison to other authors

The coefficients we derive for the FP agree reasonably well to those obtained earlier in literature, e.g., Faber et al. (1987); Djorgovski & Davis (1987); Bender et al. (1992); Guzmán et al. (1993); Saglia, Bender, & Dressler (1993); and Paper I. Most of the differences between the various results are due to the fitting method, and the selection of galaxies. The most common approach to determine the coefficients is a least squares fit with the the residuals minimized in the direction of $\log r_e$. Such a least squares fit can lead to biased results, as the velocity dispersion and mean surface brightness have significant measurement errors. Furthermore, most samples are selected by apparent magnitude and the requirement that the observed velocity dispersion is higher than 100km s$^{-1}$. This adds further biases to the coefficients.

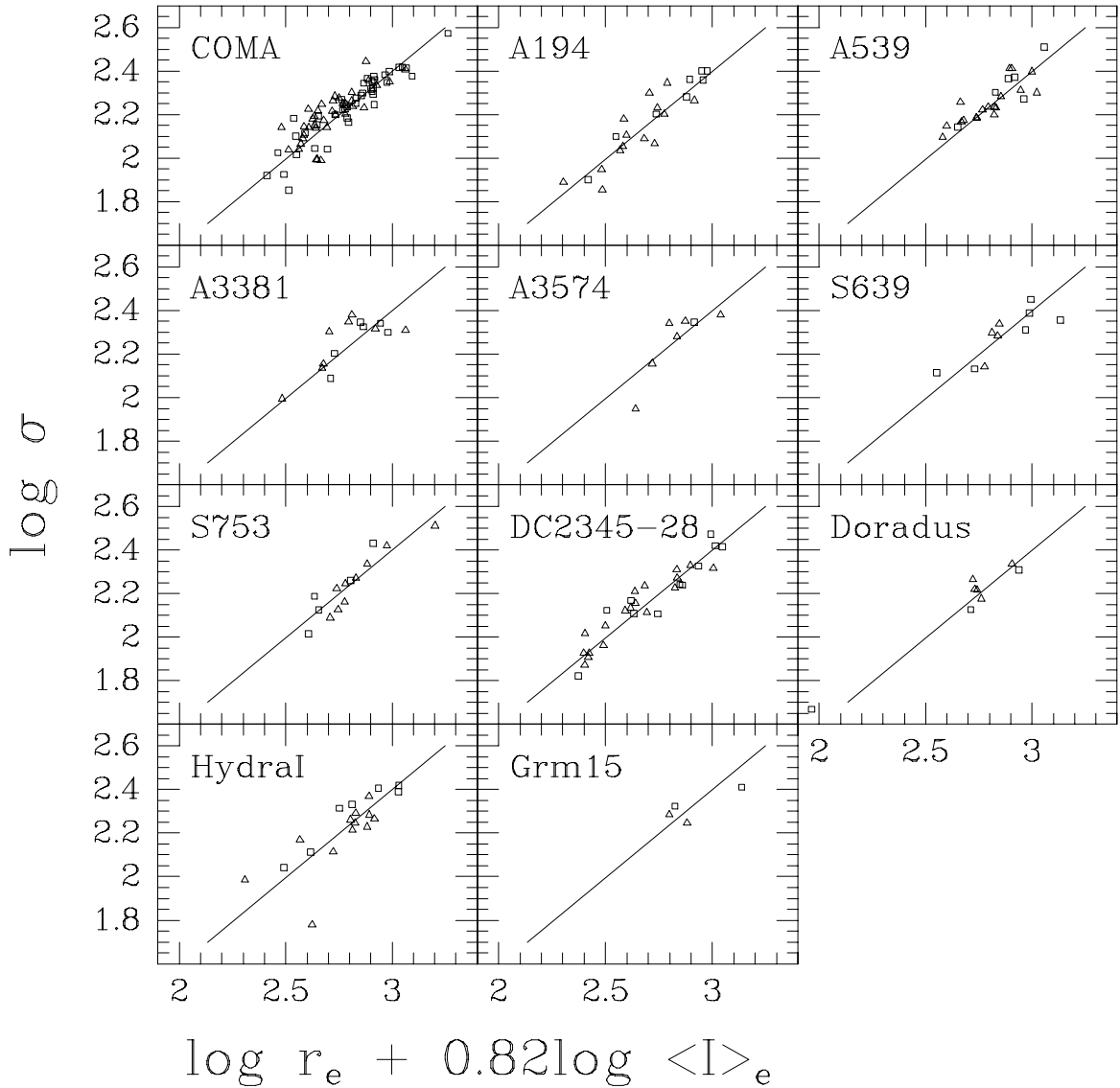

**Figure 2.** The FP in Gunn r as derived in Eq. 1 shown edge-on for all eleven clusters. Boxes – E galaxies, triangles – S0 galaxies. The cluster Grm15 has not been used for the determination of the FP.

The bias caused by measurement errors is straightforward to estimate. For uncorrelated measurement errors the estimated bias for the coefficient $c_p$ of parameter $p$ is of order $[s(p)/S(p)]^2 c_p$. Here $s(p)$ is the rms measurement error, and $S(p)$ is the rms of $p - <p>$. Typical values for $s(p)/S(p)$ are 0.15, 0.19, 0.22, and 0.11 for $\log r_e$, $\log <I>_e$, $\log \sigma$ and $\log r_e + 0.82 \log <I>_e$, respectively. A least squares fit in $\log r_e$ can therefore produce a systematic bias on the order of 5% for the coefficient $\alpha$ for $\log \sigma$. This effect can be partly circumvented by minimizing in the three directions separately, and then using some average of the results as the final fit. Further, the three determinations of the coefficients give an idea of the systematic uncertainty in the coefficients. For our sample, we obtain coefficients in the range $1.09 \leq \alpha \leq 1.39$, and $-0.87 \leq \beta \leq -0.79$. This includes the values given above, and all published results lie in this range.

The coefficients also depend on the selection criteria. For example, if the galaxies with $\log \sigma < 2.0$ are excluded the coefficient $\alpha$ rises in a systematic way. For our sample we find

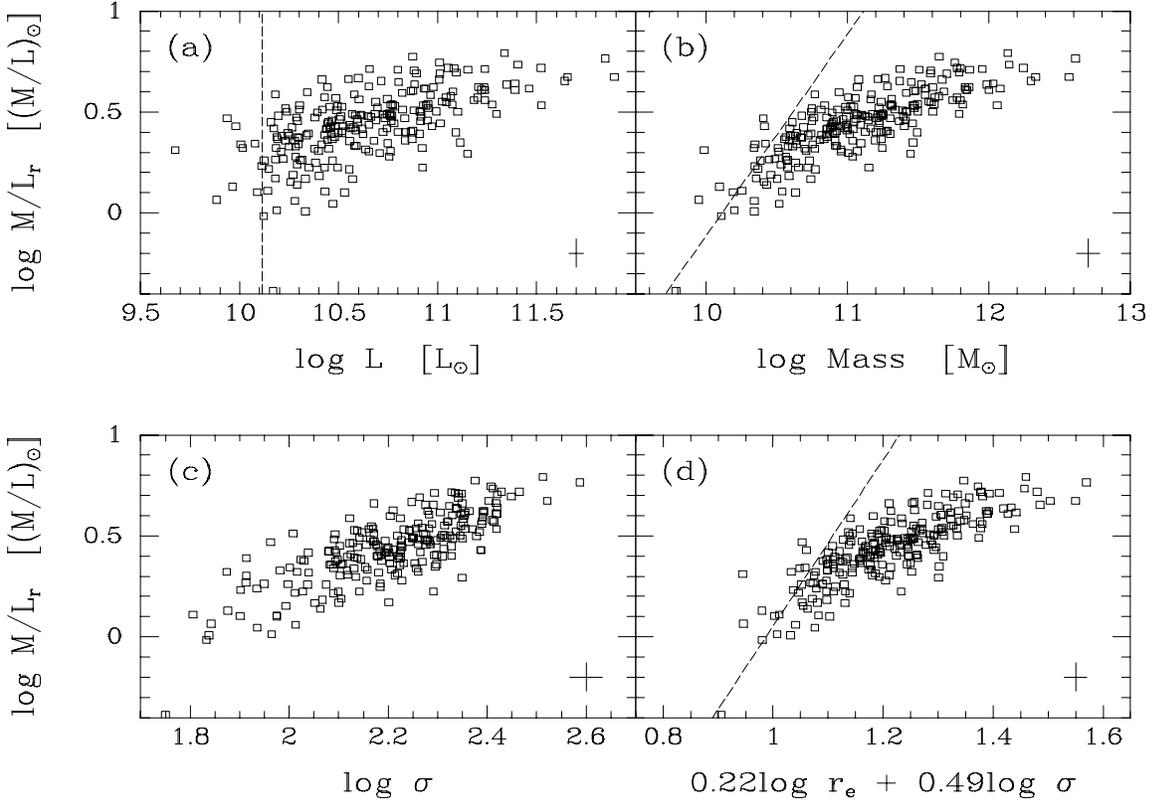

**Figure 3.** The M/L ratio as function of the luminosity, the mass, the velocity dispersion and the combination $0.22\log r_e + 0.49\log \sigma$. $\log M/L = 2\log \sigma - \log <I>_e - \log r_e +$ constant. $r_e$ is in kpc; M/L, mass and luminosity in solar units. For $H_o = 50$ km s$^{-1}$ Mpc$^{-1}$ and Mass=$5\sigma^2 r_e/G$ the constant is $-0.73$. The dashed lines on figures a, b and d show the selection effect due to a limiting magnitude of $-20^m\!\!.45$ in Gunn r. This is the limit for the Coma cluster sample.

$$\log r_e = \begin{array}{c} 1.35\log \sigma - 0.82\log <I>_e + \gamma_{cl} \\ \pm 0.05 \quad\quad \pm 0.03 \end{array} \quad (2)$$

The change of $\alpha$ may very well be due to an increased bias.

Our sample of galaxies in each cluster is not complete to a given limit, and the limiting magnitude varies from cluster to cluster. If we exclude galaxies with total absolute magnitude fainter than $-20^m\!\!.9$ in Gunn r, then the samples in all the clusters except A3574 will have the same magnitude limit. With this selection criterion the best fitting plane is $\log r_e = 1.23\log \sigma - 0.83\log <I>_e + \gamma_{cl}$. The coefficients change less then 0.01 if the magnitude limit is changed to $-21^m\!\!.5$. We conclude that our procedure is not very sensitive to the differences in limiting magnitude.

We note that the uncertainties we give in this paper are derived from a bootstrap procedure. Because of the systematic effects discussed here these uncertainties are not necessarily a true estimate of the real uncertainties.

### 3.2 The M/L ratio

As shown originally by Dressler et al. (1987b), the existence of the FP implies that the M/L ratio of galaxies is well behaved. Since the M/L ratio is a function of the stellar population, and dark matter content of galaxies, this is a truly independent and important result. Under the assumption that the galaxies have similar structure, the M/L ratio can be related to the coefficients $\alpha$, $\beta$ in the following way

$$M/L \propto r_e^{-1-1/\beta} \sigma^{2+\alpha/\beta} = r_e^{0.22} \sigma^{0.49} \propto M^{0.24} r_e^{-0.02} \quad (3)$$

$$M/L \propto <I>_e^{-(2\beta+1)/\alpha-1/2} L^{1/\alpha-1/2} = <I>_e^{0.02} L^{0.31} \quad (4)$$

Figure 3 shows the relations between the M/L ratio, and the luminosity, the mass, the velocity dispersion, and the combination $r_e^{0.22}\sigma^{0.49}$. The scatter in all relations is significant, amounting to 31% for the relation between M/L and L, and 23% between the M/L and $r_e^{0.22}\sigma^{0.49}$. The fact that the scatter is higher for the Eq. 4 than for Eq. 3 is due to a simple mathematical effect: it is easy to show that the logarithm of the residual of Eq. 4 is about 1.3 times the residual of Eq. 3.

If the structural parameters $r_e$ and $\sigma$ are given, e.g. at a given mass, then the M/L ratio is determined to 23%. The inclusion of $r_e$ is not even essential, as the relation between the M/L ratio and $\sigma$ by a direct fit, $M/L \propto \sigma^{0.86}$, has a scatter of only 25%. These results imply that the stellar population is a very strong function of the structural parameters. As an example, the M/L ratio of a stellar population depends strongly on its age and to lesser extend on its metallicity. From the stellar population models by Worthey (1994) we find $\log M/L_r \approx 0.83 \log t + 0.19[Fe/H] - 0.18$, where $t$ is the luminosity weighted mean age in Gyr of the stellar populations. The relation was derived from models with ages between 5 and 17 Gyr and [Fe/H] in the interval −0.5 to 0.5. If only differences in ages contribute to the scatter of the M/L ratio at a given mass then the allowed scatter in the ages is 0.12 dex (see also Renzini & Ciotti 1993).

Figure 3a seems to indicate that the relation between the M/L ratio and the luminosity might flatten out slightly for the most luminous galaxies. These are also the galaxies that seemed to lie slightly below the FP in the edge-on view shown in Figure 1b. The dominating selection effect (the magnitude limit for the Coma cluster sample) is shown in Figure 3a, b and d. This selection effect is (partly) the reason for the bended apparence of the relations in Figure 3b and d. We conclude again, that the FP may have a slight curvature for high luminosity galaxies, but that the effect is small.

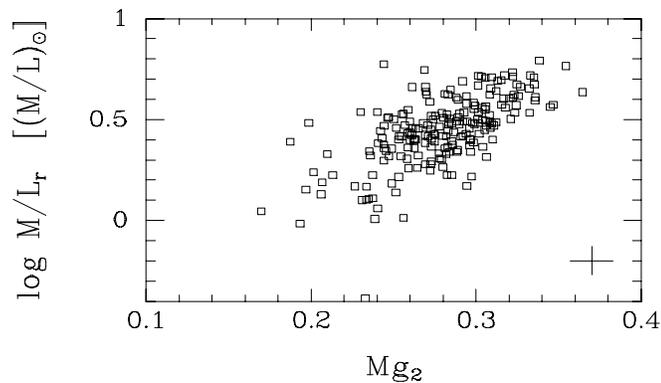

**Figure 4.** The M/L ratio as function of the $Mg_2$ line index. M/L is in solar units.

### 3.3 Correlation of the M/L ratio with stellar population

Since the M/L ratio of the galaxies is directly related to the stellar population, it is logical to ask the question whether the M/L ratio can be related directly to other observables which are functions of the stellar population. The best observable that we have is the $Mg_2$ index. It is available for 207 of the 232 galaxies. This is significantly more galaxies than we have broad band colors for. The $Mg_2$ index is a function of both the metallicity and the age of the stellar population (e.g., Worthey 1994). Figure 4 shows the M/L ratio versus the $Mg_2$ index. The parameters are highly correlated. This is not surprising, since the relation between $Mg_2$ and velocity dispersion is well established (e.g., Burstein et al. 1988). The scatter in the M/L-$Mg_2$ relation is 0.14 dex, while the scatter in the M/L-$\sigma$ relation is 0.11 dex. The difference in scatter is as expected based on the scatter in the $Mg_2$-$\sigma$ relation ($\approx 0.1$ in $\log \sigma$, cf. Sect. 6.1). Thus, whatever is the cause of the scatter in the $Mg_2$-$\sigma$ relation is most likely also the reason why the M/L-$Mg_2$ relation has higher scatter than the M/L-$\sigma$ relation.

### 3.4 Residual correlations with $\log r_e$, $\log \sigma$, $\log <I>_e$

The residuals for the FP are weakly correlated with the structural parameters $\log \sigma$ and $\log <I>_e$. If we take as the residual $\Delta FP = \log r_e - 1.24 \log \sigma + 0.82 \log <I>_e - \gamma_{cl}$, we obtain Spearman rank order correlation coefficients of $-0.015$, $-0.212$, and $0.131$ for $\log r_e$, $\log \sigma$ and $\log <I>_e$, respectively. We find similar correlations with derived quantities such as mass as with $\log \sigma$ and $\log <I>_e$. These correlations are due to the simple fact that for no value of $\alpha$ and $\beta$, are all three correlation coefficients zero. For slightly different coefficients the residuals are also correlated with absolute magnitude. It implies that a selection effect in absolute magnitude can cause a small systematic effect in the derived distance. If the selection criterion is known, then the coefficients can be chosen such that the correlation of the residual with the selection parameter disappears.

For our sample we find $\Delta FP = (0.015 \pm 0.014) \log L - 0.156$, where $\log L$ is the absolute luminosity in solar units. The clusters in our sample have at most a factor two in difference between the median absolute luminosity of the observed galaxies. From the actual differences we find that the distances relative to the distance of the Coma cluster is at most affected with $-0.5\%$ to $0.9\%$ (excl. the cluster Grm15 for which the effect is 2%). In the following relative distances derived from Eq. 1 are used. The values are given in Table 4. The distances were corrected for cosmological effects following Lynden-Bell et al. (1988). We discuss the FP as a distance determinator further in Sect. 7.

### 3.5 The thickness of the FP

The rms thickness of the FP is 0.084, measured as the rms scatter of the residuals in $\log r_e$. We tested if the distribution of the residuals is Gaussian by using a Kolmogorov-Smirnov test modified for the fact that we derive the dispersion from the data. The test gives a 56% probability that the distribution is Gaussian. The thickness is higher for galaxies with velocity dispersion below 100 km s$^{-1}$ than for galaxies with higher velocity dispersion. We find 0.125 and 0.073, respectively. Part of the difference is due to the larger measurement errors on the small velocity dispersions. However, if the estimated measurement errors are realistic this cannot explain the whole difference.

The thickness of the FP we find in this study is higher than we found earlier for a sample of 22 E galaxies in the Coma cluster (Paper I; see also Lucey, Bower, & Ellis 1991a). For the 28 E galaxies in the Coma cluster with spectroscopy from Davies et al. (1987) and velocity dispersion larger than 100 km s$^{-1}$ we find the rms scatter relative to the FP given in Eq. 1 to be 0.042. This is consistent with our previous result on the Coma cluster. The larger scatter found for the full sample of E and S0 galaxies in the ten clusters may originate partly from larger uncertainties in the velocity dispersions. However, the errors would have to be considerably

TABLE 3
THE FP IN GUNN r

| Cluster | $N_E$ | $N_{S0}$ | rms fit | rms internal | Zero point $\gamma_{cl}$ | $\Delta$(E–S0) | Individual determinations | rms |
|---|---|---|---|---|---|---|---|---|
| Coma | 42 | 37 | 0.079 | 0.070 | 0.182 | 0.035±0.018 | $1.31\log\sigma - 0.84\log\langle I\rangle_e - 0.082$ | 0.083 |
|  |  |  | 0.067 | 0.056 |  |  | ±0.07         ± 0.04 |  |
| A194 | 8 | 14 | 0.085 | 0.076 | 0.380 | −0.057±0.031 | $1.03\log\sigma - 0.83\log\langle I\rangle_e + 0.512$ | 0.073 |
|  |  |  | 0.079 | 0.071 |  |  | ±0.10         ± 0.08 |  |
| A539 | 6 | 18 | 0.071 | 0.065 | 0.113 | −0.028±0.035 | $1.32\log\sigma - 0.81\log\langle I\rangle_e - 0.177$ | 0.073 |
|  |  |  | 0.071 | 0.065 |  |  | ±0.29         ± 0.05 |  |
| A3381 | 6 | 8 | 0.105 | 0.094 | 0.016 | −0.004±0.055 | $1.22\log\sigma - 0.66\log\langle I\rangle_e - 0.309$ | 0.087 |
|  |  |  | 0.109 | 0.099 |  |  | ±0.60         ± 0.12 |  |
| A3574 | 1 | 6 | 0.102 | 0.098 | 0.393 | 0.001±0.045 | $0.59\log\sigma - 0.74\log\langle I\rangle_e + 1.294$ | 0.076 |
|  |  |  | 0.054 | 0.047 |  |  | ±2.66         ± 0.14 |  |
| S639 | 6 | 4 | 0.094 | 0.090 | 0.318 | 0.060±0.059 | $1.59\log\sigma - 0.57\log\langle I\rangle_e - 1.441$ | 0.094 |
|  |  |  | 0.094 | 0.090 |  |  | ±0.75         ± 0.14 |  |
| S753 | 5 | 9 | 0.071 | 0.065 | 0.415 | −0.010±0.044 | $1.15\log\sigma - 0.82\log\langle I\rangle_e + 0.230$ | 0.065 |
|  |  |  | 0.069 | 0.062 |  |  | ±0.26         ± 0.07 |  |
| DC2345-28 | 11 | 18 | 0.071 | 0.058 | 0.099 | 0.021±0.029 | $1.21\log\sigma - 0.82\log\langle I\rangle_e + 0.098$ | 0.070 |
|  |  |  | 0.072 | 0.063 |  |  | ±0.13         ± 0.05 |  |
| Doradus | 3 | 5 | 0.071 | 0.061 | 0.951 | 0.080±0.067 | $1.27\log\sigma - 0.74\log\langle I\rangle_e - 0.267$ | 0.055 |
|  |  |  | 0.060 | 0.054 |  |  | ±0.35         ± 0.10 |  |
| HydraI | 7 | 12 | 0.125 | 0.121 | 0.434 | −0.079±0.047 | $1.32\log\sigma - 0.87\log\langle I\rangle_e - 0.033$ | 0.135 |
|  |  |  | 0.069 | 0.064 |  |  | ±0.37         ± 0.04 |  |
| Grm15 | 2 | 2 | 0.094 | 0.088 | 0.401 | 0.011±0.115 |  |  |
|  |  |  | 0.094 | 0.088 |  |  |  |  |
| All clusters | 97 | 133 | 0.084 | 0.071 | – | 0.006±0.011 |  |  |
|  |  |  | 0.073 | 0.057 |  |  |  |  |
| E galaxies | 97 | 0 | 0.073 | 0.057 | – |  |  |  |
|  |  |  | 0.069 | 0.052 |  |  |  |  |
| S0 galaxies | 0 | 133 | 0.091 | 0.079 | – |  |  |  |
|  |  |  | 0.075 | 0.060 |  |  |  |  |

NOTE.— $N_E$ and $N_{S0}$: Number of galaxies included in the fit. "Zero point" is the constant in Eq. 1 for the individual clusters. $\Delta$(E–S0) is the difference in the median zero point for E and S0 galaxies. Grm15 was not included in the determination of the FP. The second listing of "rms fit" and "rms internal" for each cluster excludes galaxies with velocity dispersion smaller than 100 km s$^{-1}$.

TABLE 4
DISTANCES AND PECULIAR VELOCITIES

| Cluster | $N_{gal}$ | $cz_{CMB}$ [km/s] | The FP | | | The FP incl. $\langle\Delta Mg_2\rangle$ term | | | The $D_n$-$\sigma$ relation | | |
|---|---|---|---|---|---|---|---|---|---|---|---|
|  |  |  | $R_{cl}$ [km/s] | $v_{pec}$ [km/s] | Unc. [km/s] | $R_{cl}$ [km/s] | $v_{pec}$ [km/s] | Unc. [km/s] | $R_{cl}$ [km/s] | $v_{pec}$ [km/s] | Unc. [km/s] |
| Coma | 79 | 7200 | 7200 | 0 | 148 | 7200 | 0 | 151 | 7200 | 0 | 151 |
| A194 | 22 | 5038 | 4506 | 532 | 188 | 4668 | 370 | 201 | 4439 | 599 | 206 |
| A539 | 24 | 8734 | 8522 | 213 | 285 | 8894 | −159 | 311 | 8761 | −27 | 336 |
| A3381 | 14 | 11472 | 10805 | 667 | 698 | 11424 | 49 | 767 | 10373 | 1099 | 572 |
| A3574 | 7 | 4931 | 4375 | 556 | 388 | 4474 | 457 | 409 | 4248 | 683 | 463 |
| S639 | 10 | 6545 | 5250 | 1295 | 359 | 5666 | 879 | 392 | 5383 | 1162 | 333 |
| S753 | 14 | 4421 | 4141 | 279 | 182 | 4303 | 118 | 194 | 4047 | 374 | 228 |
| DC2345-28 | 29 | 8482 | 8773 | −291 | 265 | 8719 | −238 | 309 | 8250 | 232 | 278 |
| Doradus | 8 | 1104 | 1185 | −81 | 68 | 1248 | −144 | 78 | 1257 | −153 | 82 |
| HydraI | 19 | 4050 | 3963 | 87 | 261 | 4001 | 49 | 267 | 3953 | 97 | 263 |
| Grm15 | 4 | 4575 | 4286 | 289 | 465 | 4307 | 267 | 494 | 4218 | 356 | 412 |

NOTE.— $N_{gal}$: Number of galaxies used for the determinations. $R_{cl}$: Relative distance derived from the FP or the $D_n$-$\sigma$ relation, given in velocity units. $v_{pec}$: Peculiar velocity ($\equiv cz_{CMB} - R_{cl}$). The Coma cluster is assumed to have $v_{pec}$=0. Uncertainties of distances based on the FP or the $D_n$-$\sigma$ relation are derived from the rms scatter for each cluster relative to the relation. For the FP incl. the $\langle\Delta Mg_2\rangle$ term the uncertainty on the $\langle\Delta Mg_2\rangle$ term was added in quadrature.

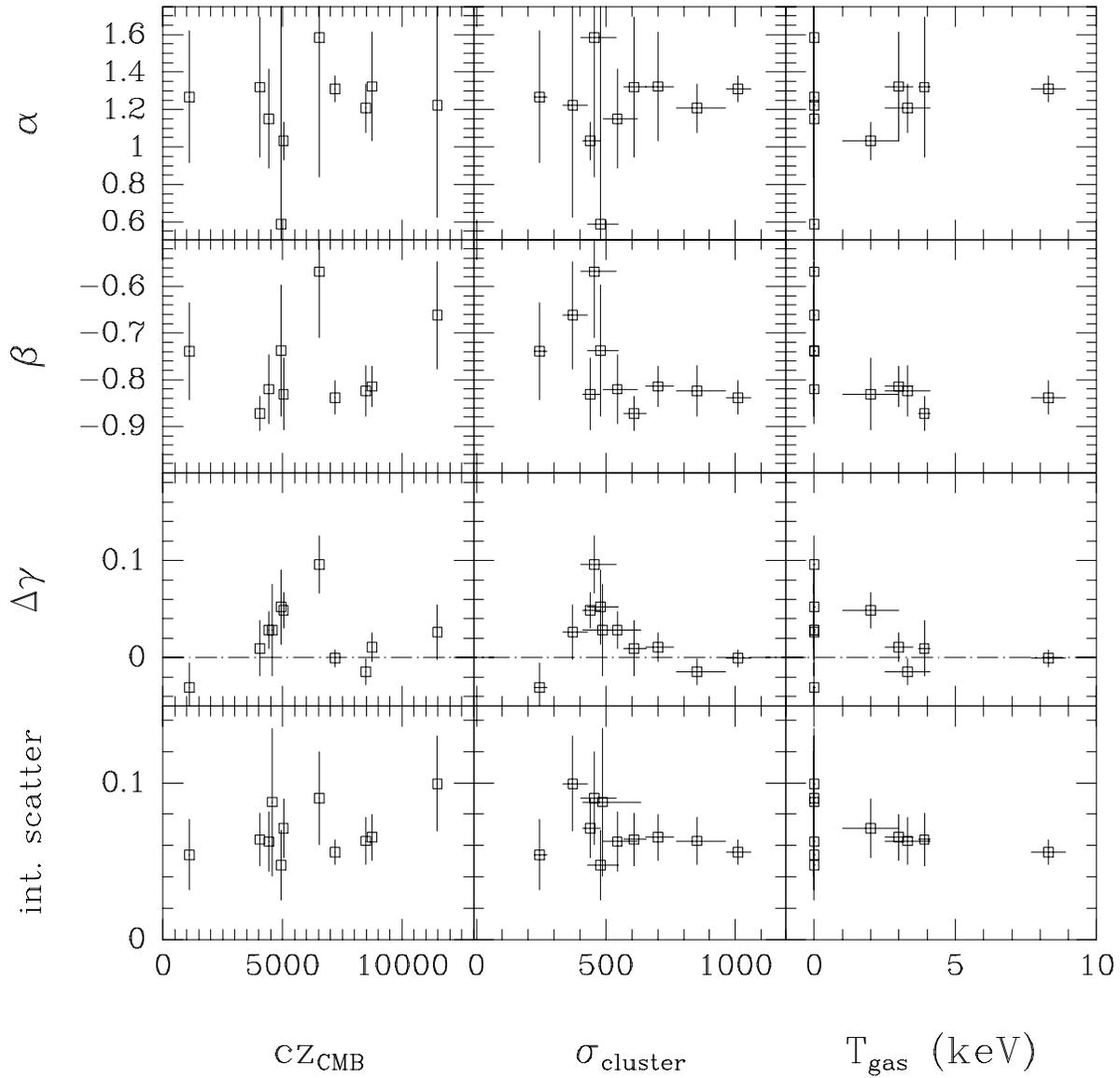

**Figure 5.** The coefficients ($\alpha, \beta$) for the individual determinations of the FP, and $\Delta\gamma \equiv \log(v_{\rm pec}/R_{\rm cl} + 1)$ versus cluster properties. $\sigma_{\rm cluster}$ is the velocity dispersion of the cluster. $T_{\rm gas}$ is the intracluster gas temperature derived from X-ray observations. In the bottom panels the intrinsic scatter relative to the common FP (Eq. 1) is plotted versus the cluster properties. Galaxies with velocity dispersion below $100 {\rm km\,s^{-1}}$ are omitted in the estimates of the intrinsic scatter. No significant correlations with the cluster properties are found.

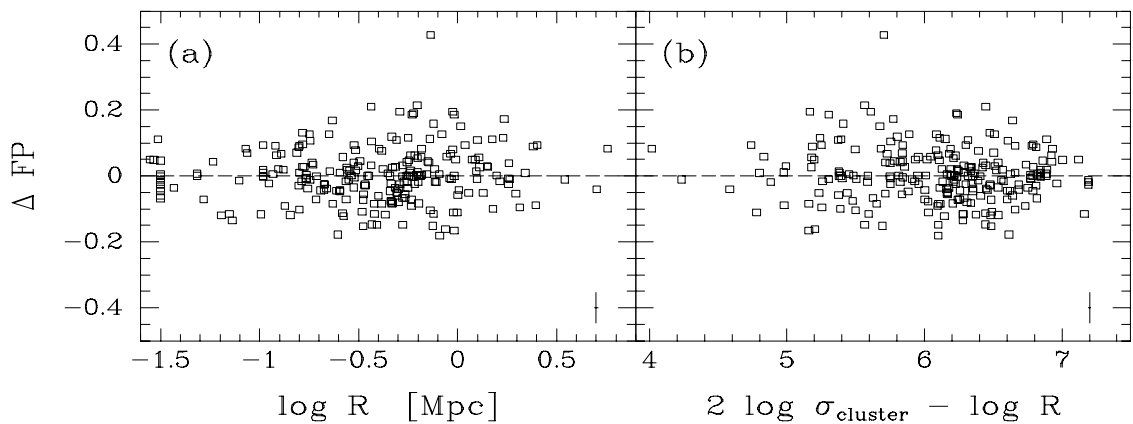

**Figure 6.** Residuals for the FP versus cluster center distance and the projected cluster surface density $\sigma^2_{\mathrm{cluster}}/R$. The residuals are calculated as $\Delta \mathrm{FP} = \log r_e - \mathrm{fit}$. $\sigma_{\mathrm{cluster}}$ is in km s$^{-1}$, $R$ in Mpc.

larger than the formal errors to cause the increased thickness. Real effects from galaxies with strong disks seen close to edge-on and galaxies with larger variations in the stellar populations than the E galaxies in the Coma cluster must also contribute.

The scatter around the FP may be a function of galaxy properties. In Paper II we showed that E galaxies brighter than $-23\overset{m}{.}1$ in Gunn r have no disks, whereas E and S0 galaxies fainter than $-23\overset{m}{.}1$ have a large spread in bulge to disk ratios. When we make a similar cut for the galaxies in this sample, we find the rms scatter to be 0.047 and 0.086 for the bright and the faint galaxies, respectively. The difference is significant. It may be due to the presence of a disk in some of the lower luminosity galaxies. Also larger variations in the stellar populations for the lower luminosity galaxies may contribute, since the range of the Mg$_2$ index is larger for these galaxies than for the bright galaxies. The scatter around the FP for the bright galaxies is not larger than expected from measurement errors.

The mean intrinsic scatter relative to the FP (Eq. 1) was derived by subtraction of the measurement uncertainties in quadrature from the rms scatter of the fit. The correlation of the errors in $\log r_e$ and $\log <I>_e$ has been taken into account. We find the intrinsic scatter of the FP to be 0.070 in $\log r_e$. The intrinsic scatter for galaxies with velocity dispersion larger than 100 km s$^{-1}$ is 0.057. Because the thickness of the FP is higher than expected from measurement errors, we have to conclude that either unknown observational errors exist, or that there is a significant source of internal scatter. This source of scatter is a great concern, because without more information we cannot exclude that an undetermined physical effect is responsible and that such an effect can cause systematic biases in the distances derived to galaxies and clusters. In the following we therefore investigate in detail if the residuals are correlated with any other observables.

## 4 THE FP AS A FUNCTION OF CLUSTER ENVIRONMENT

In this section we test whether the internal scatter for the FP is related to the the cluster environment. First we determined the coefficients for the FP for each cluster separately, to test if the coefficients depend on cluster properties. The individual fits were determined in the same way as the determination of the common FP, e.g. as orthogonal fits. The individual fits are given in Table 3. Individual values of the rms scatter and the intrinsic scatter relative to the common FP (Eq. 1) are also listed in the table.

In Fig. 5 the coefficients for the individual determinations and the intrinsic scatter relative to the common FP are plotted versus the radial velocity and the velocity dispersion of the clusters, and versus the temperature of the intracluster gas as derived from X-ray observations. Also $\Delta \gamma \equiv \log(v_{\mathrm{pec}}/\mathrm{R}_{\mathrm{cl}} + 1)$ is plotted versus the cluster parameters. $\mathrm{R}_{\mathrm{cl}}$ is the distance of the clusters in velocity units. If FP is universal and there are no peculiar velocities then $\Delta \gamma = 0$.

None of the parameters correlate with radial velocity. Furthermore, there is no significant dependence on $T_{\mathrm{gas}}$, Abell richness, or cluster velocity dispersion. It is remarkable that neither the coefficients nor the intrinsic scatter show any dependence on the cluster properties. Variations of $\alpha$ of the order $\pm 10\%$ cannot be ruled out by the data. The interval spanned by the individual determinations of $\alpha$, roughly $1.2 \pm 0.2$, translates into $M/L \propto L^{0.35 \pm 0.15}$. It will be valuable to derive the FP for larger samples of homogeneously selected cluster galaxies to set further limits on the possible variation of $\alpha$, and to establish the proportionality $M/L \propto L^a$ with better accuracy.

We conclude that the data do not show any significant evidence that the coefficients of the FP or the intrinsic scatter depend on any of the global cluster properties. Also, the individual determinations of the FP are not significantly different from the common FP (Eq. 1).

Figure 6 shows the residuals for the FP versus the cluster center distance on a metric scale and versus an estimate of the projected cluster surface density, $\rho_{\rm cluster}$. $\rho_{\rm cluster}$ has been derived from the approximate expression

$$\rho_{\rm cluster} \propto {\rm Mass}/R^2 \propto \sigma_{\rm cluster}^2/R \qquad (5)$$

The center of each cluster has been taken to be the brightest cluster member, or in cases where there are two comparable bright members then the mean position of these. There is no dependence of the cluster center distance or $\rho_{\rm cluster}$, neither of the intrinsic scatter nor of the zero point. The zero point is stable within $0.014\pm0.012$ over the range of cluster center distances and projected densities observed. This is equivalent to $3\%\pm3\%$ if the FP is used for distance determinations. The sample of 66 galaxies in the Coma cluster observed by Lucey et al. (1991b) has a comparable stability of the zero point for the FP. This sample spans a factor 150 in projected surface density of the galaxies, and cluster center distances as large as 6 Mpc h$^{-1}$.

probability that two samples were drawn from distributions with the same mean value (van der Waerden 1969). We find a probability of 14% that this is the case for the residuals for the E and the S0 galaxies. The median difference in $\Delta$FP is also very small, only $0.006 \pm 0.011$. This is equivalent to $1.4\%\pm2.5\%$ difference on distance determinations. Further, our determinations of the FP in other passbands show very small zero point differences between E and S0 galaxies, cf. Table 5.

Our result differs from the work by Saglia et al. (1993). They found an offset of 0.043 between S0 galaxies and ellipticals in the Coma cluster. The difference originates from the coefficients for the FP these authors adopted and from the sample selection. If we use their coefficients ($\alpha = 1.07$ $\beta = -0.87$) we find a difference for our full sample of 0.022. If we use their coefficients for only our Coma cluster sample we find a difference of 0.068.

## 5 THE FP AS A FUNCTION OF GALAXY MORPHOLOGY

In Paper II we found that most E and S0 galaxies in Coma have a very wide distribution of the relative disk luminosity, $L_{\rm D}/L_{\rm tot}$, spanning the whole interval from zero to one. Strong disks are expected to leave their signatures on the ellipticity, isophotal shapes and, possibly, the derived M/L ratios. Here we investigate if such effects show up in the residuals for the FP.

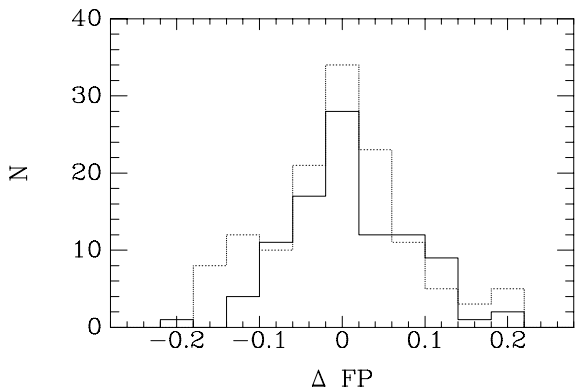

**Figure 7.** Histogram of the residuals for the FP (Eq. 1). Solid line – E galaxies. Dotted line – S0 galaxies.

### 5.1 Galaxy type

The FP was derived for the E and S0 galaxies, separately. The coefficients were not significantly different from the coefficients given in Eq. 1. Histograms of the residuals with respect to the common FP are shown in Fig. 7. The distributions of the residuals are very similar for the E and the S0 galaxies. We have tested this with a Mann-Whitney test. The Mann-Whitney test is a rank order test, which gives the

### 5.2 Ellipticity, $<c_4>$, and disk luminosity

The addition of a disk to a spheroidal galaxy will influence all observed parameters, and can easily cause offsets from the FP. The direction and size of the effect will depend on the viewing angle, as the geometry of the disk is very different from the geometry of the bulge/spheroid. We have constructed axisymmetric models of galaxies consisting of a disk and a bulge, and with relative disk luminosities, $L_{\rm D}/L_{\rm tot}$, between zero and one. Pseudo photometry was made for these models. The disk was assumed to have an exponential luminosity profile while the bulge was assumed to have an $r^{1/4}$ profile. Both components were assumed to be oblate. The intrinsic ellipticities were 0.85 and 0.3 for the disk and the bulge, respectively. The seeing effects were taken into account. Kinematic models were produced under the assumption that the distribution function is a function only of energy $E$ and angular momentum $L_z$ around the z-axis. We have used the programs by van der Marel (1991) for making the kinematic models.

The models predict various relations between the residuals for the FP and the ellipticity, the isophotal shape $<c_4>$ and $L_{\rm D}/L_{\rm tot}$. Figure 8 shows the data together with the models. The ellipticity and $<c_4>$ were used to estimate the relative disk luminosities for galaxies with inclination larger than $\approx 60°$. This technique is described in Paper II. The data do not to follow the trends indicated by the models closely, except that galaxies with high ellipticity have positive residuals. Spearman rank order tests show that the residuals are weakly correlated with both ellipticity and $<c_4>$. These correlations are caused by the 20 galaxies with ellipticity larger than 0.6. The residuals and $L_{\rm D}/L_{\rm tot}$ are not correlated. According to the models a strong correlation with $L_{\rm D}/L_{\rm tot}$ was expected. None of the parameters can help to reduce the scatter in the FP in a significant way. It is somewhat surprising that the data do not follow the model predictions, since the analysis of the models took all observational procedures into account. More detailed data will be needed to determine how variations in $L_{\rm D}/L_{\rm tot}$ affect the FP.

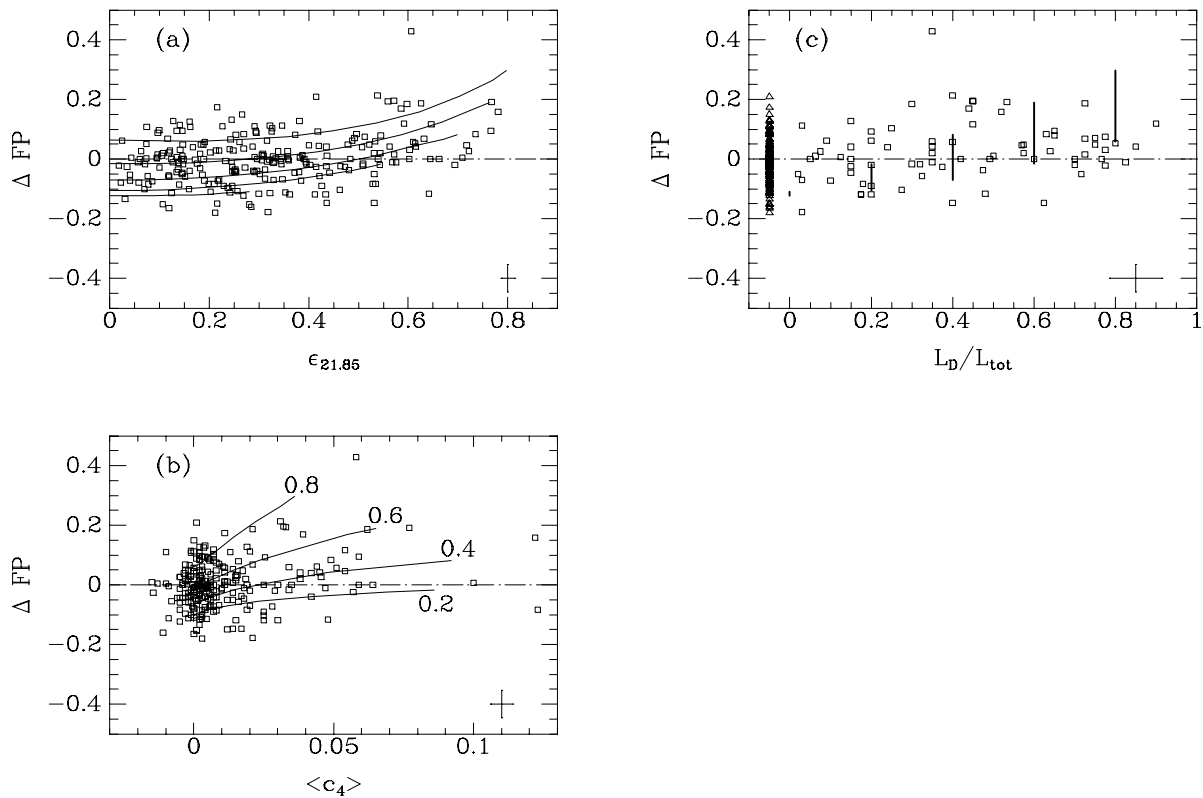

**Figure 8.** The residuals for the FP (Eq. 1) versus (a) the ellipticity $\epsilon_{21.85}$, (b) the parameter $<c_4>$, and (c) the relative disk luminosity $L_D/L_{tot}$. The residuals are calculated as $\Delta FP = \log r_e -$ fit. $\epsilon_{21.85}$ is the ellipticity at a local surface brightness of 21.85 mag arcsec$^{-2}$ in Gunn r. $<c_4>$ is the intensity weighted mean of the coefficient for the $\cos 4\theta$ term in a Fourier expansion of the deviations of the isophotes from ellipses. $<c_4> > 0$ means the galaxy is disky, $<c_4> < 0$ that it is boxy. $L_D/L_{tot}$ has been estimated from $\epsilon_{21.85}$ and $<c_4>$ (see Paper II). The triangles on (c) at $L_D/L_{tot} = -0.05$ are galaxies for which a determination of $L_D/L_{tot}$ was not possible. Typical measurement uncertainties are given in the lower right corners. The expected residuals based on simple models are overplotted. The models on (a) have (from the bottom) $L_D/L_{tot} = 0.0, 0.2, 0.4, 0.6,$ and 0.8. The models on (b) are labeled with $L_D/L_{tot}$.

## 6 THE RELATION BETWEEN THE FP AND THE STELLAR POPULATIONS

The existence of the FP implies that the M/L ratios of E and S0 galaxies are very regular, as described above. Hence the stellar populations of the galaxies must be very regular, as any differences in age or metallicity would be reflected directly in the M/L ratio through a change in the luminosity.

The underlying cause for the correlation of the M/L ratio with structural parameters is not well known. Both metallicity effects and age variations have been suggested as the main cause (Faber et al. 1987, 1995).

Here we analyze the relation between the FP and other indicators of the stellar population. These other indicators are the $Mg_2$ line index and the broad band colors. The $Mg_2$-$\sigma$ relation can be used as an indicator of small population differences between galaxies with the same velocity dispersion. If galaxies have undergone a starburst, are slightly younger or have a lower metallicity than the other galaxies at that velocity dispersion, they will have negative residuals. Such effects will also change the M/L ratios of the galaxies, and will therefore produce residuals in the FP (see also Gregg 1992, 1995; Guzmán et al. 1992). The FP with the $Mg_2$ index substituted for the velocity dispersion can be used in a similar way to investigate the age and metallicity effects. New results by Worthey, Trager & Faber (1995) indicate that age and metallicity variations are linked so more metal rich galaxies look younger. This may maintain the thinness of the FP.

### 6.1 $Mg_2$ relations

The $Mg_2$ line index has been measured for 207 of the galaxies in our sample, and is the best available indicator of the stellar population. It is well known that the $Mg_2$ index correlates strongly with the velocity dispersion (e.g., Burstein et al. 1988; Bender et al. 1993). Figure 9 shows $Mg_2$ versus velocity dispersion.

We have derived the $Mg_2$-$\sigma$ relation as the orthogonal fit with the sum of the absolute residuals minimized. The best fitting relation is

$$Mg_2 = \begin{array}{c} 0.196 \log \sigma - 0.155 \\ \pm 0.016 \end{array} \qquad (6)$$

The rms scatter of the residuals in $Mg_2$ is 0.023. The scatter is higher for the galaxies with velocity dispersion below 100km s$^{-1}$ than for the galaxies with higher velocity disper-sion, 0.036 and 0.022 respectively. The difference is (partly) due to larger measurement errors for the low velocity dispersion galaxies. The coefficient derived here agrees with determinations by Burstein et al. and Bender et al.

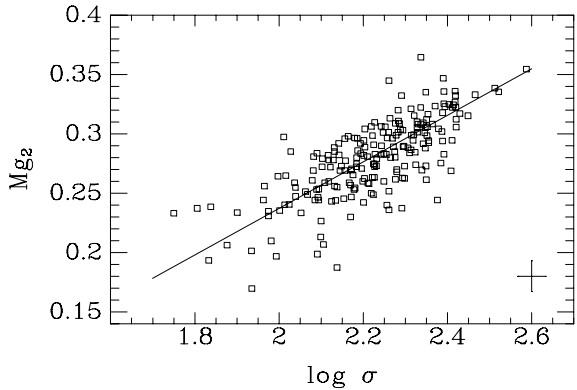

**Figure 9.** The $Mg_2$ line index versus the velocity dispersion. The typical measurement uncertainties are given in the lower right corner. The line is the relation derived in Eq. 6.

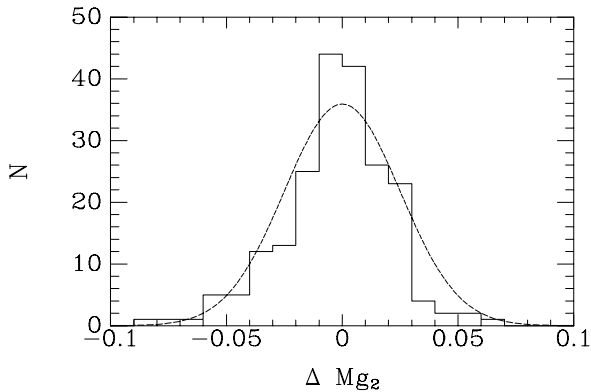

**Figure 10.** Histogram of the residuals for the $Mg_2$-$\sigma$ relation. A Gaussian distribution with a standard deviation of 0.023 is overplotted.

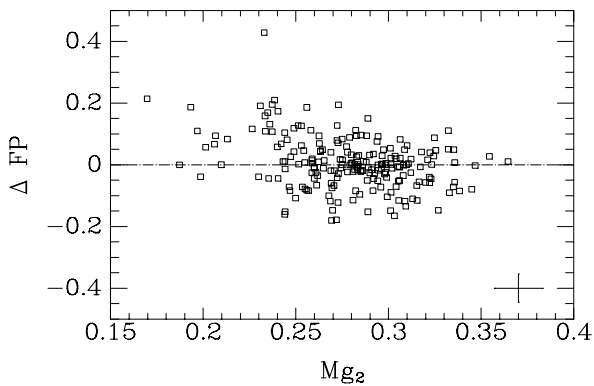

**Figure 11.** The residuals for the FP (Eq. 1) versus the line index $Mg_2$. The residuals are calculated as $\Delta FP = \log r_e - $ fit. Symbols as in Fig. 2. The residuals are weakly correlated with $Mg_2$. Part of the correlation may be due to "left over" correlation with the velocity dispersion, see text.

If we use spectroscopic parameters corrected to a circular aperture with diameter $r_e/4$ we find the best fitting $Mg_2$-$\sigma$ relation to be $Mg_2 = 0.156 \log \sigma - 0.057$. However, the aperture correction which for both parameters depends on the effective radius (cf. Paper IV) causes the residuals for this relation to be correlated with $\log r_e$. In the following we therefore use the relation given in Eq. 6.

A histogram of the residuals for Eq. 6 is shown in Figure 10. A Gaussian with the same dispersion and zero mean has been overplotted. A modified Kolmogorov–Smirnov test gives 15% probability that the residuals have a Gaussian distribution. The distribution has some tail of galaxies with negative residuals as also found by Burstein et al. for an all sky sample. However, the presence of this tail is sensitive to the adopted coefficient for the relation. For the orthogonal fit, or a fit with residuals minimized in $Mg_2$ only, the residuals correlate with $Mg_2$. The distribution of the residuals will therefore resemble that of $Mg_2$ itself. If the $Mg_2$-$\sigma$ relation is derived as the bi-sector of two fits where the residuals are minimized in each direction independently we find a coefficient of 0.254±0.017. In this case the distribution of the residuals does not show a tail of negative residuals.

The residuals for the $Mg_2$-$\sigma$ relation, $\Delta Mg_2$, do not correlate significantly with other structural parameters ($r_e$, $<I>_e$, or absolute luminosity). If included directly in the relation none of these parameters have significant coefficients.

The residuals for the FP (Eq. 1), $\Delta FP$, correlate weakly with the $Mg_2$ line index, see Figure 11. Part of the apparent correlation may be due to "left over" correlation with $\log \sigma$, cf. Sect. 3.4. Further, a direct least squares fit with the residuals minimized in $\log r_e$ only, does not show a $Mg_2$ or $\Delta Mg_2$ term to be significant. We note, however, that part of a real effect may have been hidden by the use of free zero points for the clusters in the determination of the FP. If we instead assume that the peculiar velocities are negligible and use the relative distances of the clusters given by their radial velocities relative to the CMB system, a direct least squares fit shows the $\Delta Mg_2$ term to be significant on the 1.5$\sigma$ level:

$$\log r_e = \begin{array}{c} 1.05 \log \sigma - 0.78 \log <I>_e - 0.40 \Delta Mg_2 + \gamma_{cl} \\ \pm 0.04 \quad\quad \pm 0.02 \quad\quad\quad \pm 0.26 \end{array} \quad (7)$$

The rms scatter for this relation is not smaller than for the original form of the FP fitted with fixed distances; 0.085 in $\log r_e$. Thus, the significance of the $\Delta Mg_2$ term seems to be rather small, and absolute value of the coefficient not larger than one.

Another way of investigating the importance of the stellar populations on FP-type relations is to substitute the $Mg_2$ line index for the velocity dispersion. We have derived this relation for the 207 galaxies for which $Mg_2$ is available. The orthogonal fit is

$$\log r_e = \begin{array}{c} 7.01 Mg_2 - 0.79 \log <I>_e + \gamma_{cl} \\ \pm 0.65 \quad \pm 0.04 \end{array} \quad (8)$$

with an rms scatter of 0.189 in $\log r_e$. The zero points for the clusters were defined from the relative distances derived from the FP (Eq. 1). It requires a factor 2–2.5 larger mea-surement uncertainty on Mg$_2$ than estimated from the data (Paper IV; Davies et al. 1987) if the relation is required to have the same intrinsic scatter as the FP. Thus, we conclude that real variations in the stellar populations that affect Mg$_2$ contribute significantly to the scatter in this relation.

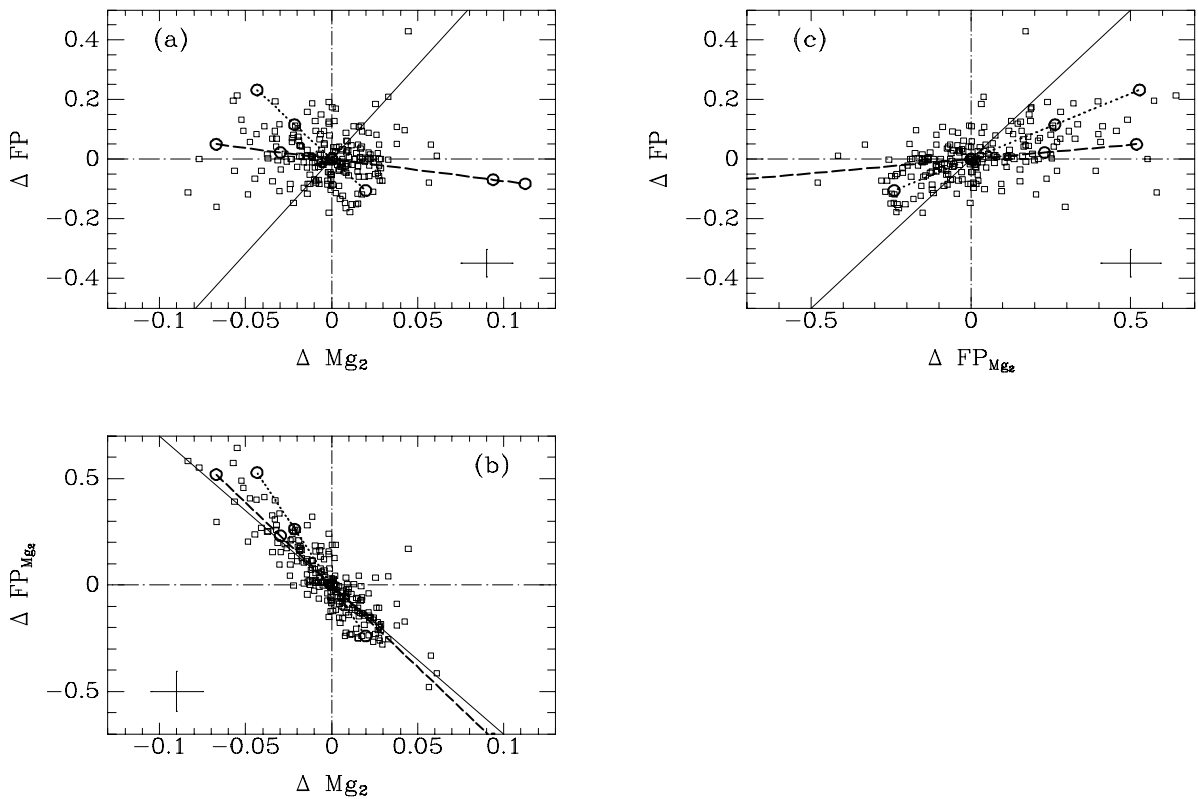

**Figure 12.** $\delta$-$\delta$ diagrams of residuals versus residuals for all derived relations. $\Delta$FP are the residuals for the FP in Gunn r (Eq. 1). $\Delta$FP$_{Mg_2}$ are the residuals for the FP where Mg$_2$ is substituted for $\log \sigma$ (Eq. 8). $\Delta$Mg$_2$ are the residuals for the Mg$_2$-$\sigma$ relation (Eq. 6). Typical error bars are given in the panels. The solid lines are the expected correlations due to the measurement errors on the parameters in common.
Predictions of residuals based on stellar populations models by Worthey (1994) are overplotted. Dotted lines – variation in age, only. Dashed lines – variation in [Fe/H], only. The predictions have arbitrarily been set to zero residuals at [Fe/H]=0 and an age of 12Gyr. The circles along the model lines mark ages of 5, 8, 12, and 17 Gyr, and [Fe/H]=–0.5, –0.25, 0.0, 0.25, and 0.5. Low ages and low [Fe/H] give positive $\Delta$FP and negative $\Delta$Mg$_2$. Not all values of [Fe/H] are shown on all plots.

Figure 12 shows the residuals for all investigated relations plotted versus each other. The solid lines on Fig. 12 show the expected correlations caused by measurement errors on the parameters in common. The data on Fig. 12b-c show correlations, which are close to the expectations, but the residuals are larger than what can be accounted for by the measurement errors. This has previously been noted by Dressler et al. (1987b) for residuals relative to the $D_n$-$\sigma$ relation and the $D_n$-Mg$_2$ relation. The residuals for the Mg$_2$-$\sigma$ relation are not correlated with the residuals for the FP, see Fig. 12a. This was also found by Bender et al. (1993). If the correlations seen on Fig. 12b-c are due to measurement errors, only, we would also expect a significant correlation between the residuals for the Mg$_2$-$\sigma$ relation and the residuals for the FP.

We can use models of the stellar populations to investigate which correlations between the residuals we may expect from variations in [Fe/H] and/or age. To do this we assume that only Mg$_2$ and the M/L ratio (through a change in luminosity) are affected by variations in [Fe/H] and age. We note, however, that the predictions will be quite model dependent. Even in the simplest case of age variation between populations, the dependence of the M/L ratio on age is a function of the IMF (e.g., Tinsley 1972).

We used the stellar population models by Worthey (1994) to estimate the expected correlations. For a simple population with a single age and single metallicity we expect $\Delta \log M/L = 6.5 \Delta \mathrm{Mg}_2$ for small changes in age, and $\Delta \log M/L = 0.9 \Delta \mathrm{Mg}_2$ for metallicity differences. The two relations between $\Delta \log M/L$ and $\Delta \mathrm{Mg}_2$ have been derived by fitting linear relations to $\Delta \log M/L$ as function of $\Delta \mathrm{Mg}_2$ at constant [Fe/H] and age, respectively. They are valid for [Fe/H] in the interval $-0.5$ to $0.5$ and ages of 5–17Gyr.

The coefficients change if we allow for more complex populations consisting of multiple populations. For example, if we assume that 10% of the mass was involved in a starburst then $\Delta \log M/L = 5.7 \Delta \mathrm{Mg}_2$. We have assumed the rest of the galaxy to be 12Gyr old and have [Fe/H]=0.

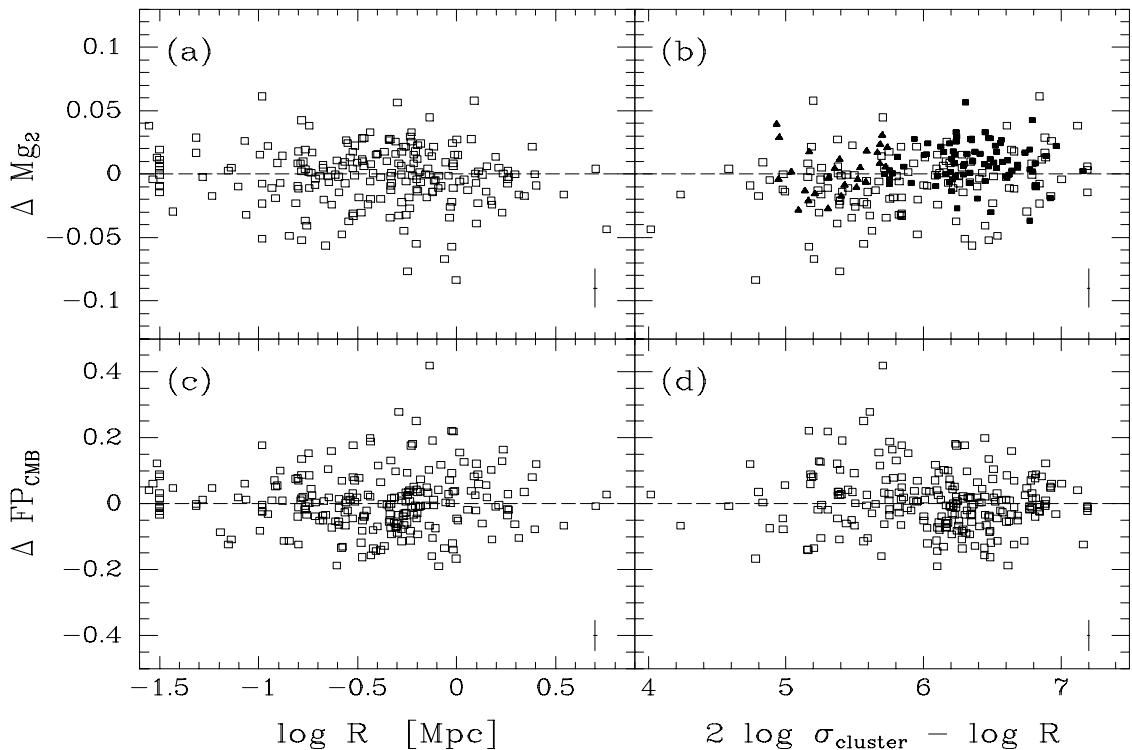

**Figure 13.** (a)–(b) Residuals for the $Mg_2$-$\sigma$ relation versus cluster center distance and an estimate of the local surface cluster density, respectively. The residuals are calculated as $\Delta Mg_2 = Mg_2 -$ fit. In (b) our sample in the Coma cluster is plotted as solid boxes and the part of the Coma sample by Guzmán et al. (1992) that is not included in our sample is shown as solid triangles. (c)–(d) Under the assumption that there are no peculiar velocities these panels show the residuals for the FP versus the same parameters. The residuals are calculated as $\Delta FP = \log r_e -$ fit.

The predictions are highly model dependent, and it will be very hard to interpret the results uniquely.

The model predictions for age and metallicity variations are overplotted on Fig. 12. The residuals for the FPs relate to $\Delta \log M/L$ and $\Delta Mg_2$ in the following way: $\Delta FP = -0.82 \Delta \log M/L$, $\Delta FP_{Mg_2} = -7.01 \Delta Mg_2 - 0.79 \Delta \log M/L$. For $\Delta FP_{Mg_2}$ versus $\Delta FP$ and versus $\Delta Mg_2$ (Fig. 12b-c) the model predictions are close to the correlations which would arise from the measurement errors. Thus, variations in [Fe/H] and age would not change the slope of these correlations very much, but just make it seem as if we had underestimated the measurement errors. For $\Delta FP$ versus $\Delta Mg_2$, however, the model predictions are at rather large angles with the correlations expected from the measurement errors. Variations in [Fe/H] and age will in this diagram weaken any correlation due to measurement errors. If the variations in the stellar populations are big enough a correlation due to measurement errors will become insignificant. This seems to be the case.

All three derived relations are affected by both age and metallicity variations. It is not possible based on the present data to conclude which effect is dominating, if any. Other line indices are needed to resolve this issue.

### 6.2 The $Mg_2$ relations as function of cluster environment

Guzmán et al. (1992) have noted that there are small systematic differences in the $Mg_2$-$\sigma$ relation and the $D_n$-$Mg_2$ relation between galaxies in the inner and outer parts of Coma. These authors found the residuals for the $Mg_2$-$\sigma$ relation to depend on the cluster center distance. This result led us to investigate whether we could find any such differences between our clusters, and within our clusters. Figure 13a-b show the residuals for the $Mg_2$-$\sigma$ relations against cluster center distance $R$, and the projected cluster surface density, $\rho_{cluster} = \sigma^2_{cluster}/R$. The residuals show no significant correlation with $R$, while there is a strong correlation with $\rho_{cluster}$. A least squares fit gives

$$\Delta Mg_2 = (0.012 \pm 0.003) \log \rho_{cluster} - 0.075 \qquad (9)$$

The median zero points for the $Mg_2$-$\sigma$ relation, $<\Delta Mg_2>$, and for the relation in Eq. 8, $<\Delta FP_{Mg_2}>$, for each cluster are plotted against cluster properties in Figure 14. The significance of the relations was tested with a Kendall's $\tau$ correlation test. (The Spearman rank order test becomes unreliable for very small samples.) The correlation between the zero points for the $Mg_2$-$\sigma$ relation and the cluster velocity dispersion is significant on the 96% level.

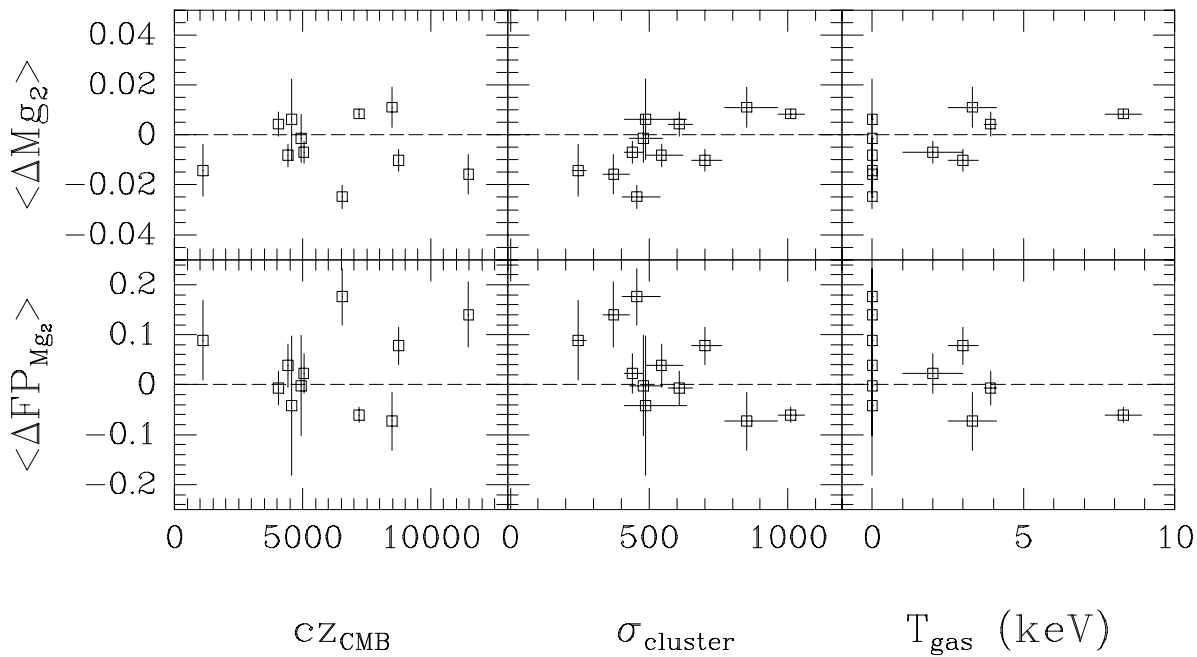

**Figure 14.** Median zero points for the $Mg_2$-$\sigma$ relation and the FP with $Mg_2$ substituted for the velocity dispersions plotted versus cluster properties. $<\Delta Mg_2>$ and $<\Delta FP_{Mg_2}>$ are correlated with $\sigma_{cluster}$.

The zero points may also be correlated with the X-ray gas temperature. The zero points $<\Delta FP_{Mg_2}>$ show the same correlations.

The correlations of $\Delta Mg_2$ with $\rho_{cluster}$ and $<\Delta Mg_2>$ with $\sigma_{cluster}$ show that galaxies in clusters with lower cluster dispersion (and lower density) have slightly lower $Mg_2$. These galaxies are on average younger and/or have lower metallicity. The implication is, that the M/L ratio of the galaxies in these clusters may be different, and that the zero points for the FP are different for different clusters. This prediction is difficult to test directly, since we do not have independent distance measurements to these clusters. Nevertheless, we can attempt to measure an effect by correlating the residual from the FP with projected cluster surface density, under the assumption that the peculiar motions of the clusters are negligible. Figure 13c-d shows the result. We do not find any significant correlations; a Spearman rank order test gives 18% probability that there is no correlation between the residuals and $\rho_{cluster}$. However, a small zero point difference may be present between galaxies at $\log \rho_{cluster} \leq 6$ and galaxies in higher density environments. The median difference in 0.035±0.013. Further, if we include $\log \rho_{cluster}$ in a direct least squares fit of the FP with the zero points left free, the significance of the term is on the $2\sigma$ level. Thus, there are weak indications that the zero point differences in the $Mg_2$-$\sigma$ relation are also reflected in the FP.

The size of the offset in $Mg_2$ we find for our clusters is of the order 0.02. If the whole effect is caused by age differences the the expected effect in the zero point of the FP might be quite high. Based on Worthey's models (1994) an offset of 0.02 in $Mg_2$ translates to 0.13dex in the M/L ratio or 0.11 in $\Delta FP$. This is much more than we see for any of the clusters, except for S639. On the other hand if the offset in $Mg_2$ is due to metallicity differences only, then the offset in $\Delta FP$ would be 0.015. This is consistent with our data. If we estimate the expected offset in the M/L ratio from the relation between the M/L ratio and $Mg_2$ (Fig. 4) we get 0.06dex, or $\Delta FP = 0.05$. This is also consistent with our data. Note that the effect Guzmán et al. (1992) found is similar: an offset in the $Mg_2$-sigma relation within Coma, while the offset in the FP for this sample is very small (Lucey et al. 1991b). For reference their data are included on Figure 13b.

The cause of the correlation between the residuals for the $Mg_2$-$\sigma$ relation and the projected cluster surface density, as well as the correlation between the median zero points and the cluster velocity dispersions, remains unresolved. It would clearly be highly valuable to obtain a very homogeneous data set on cluster galaxies in many different clusters. The current data set is from many different sources. Although the comparisons do not indicate large systematic problems, it seems that the final determination of the effects can only be done on the basis of a more homogeneous data set.

### 6.3 The FP in other passbands

We have photometric parameters for part of our sample in Gunn g, Johnson B, and Johnson U. We have derived the FP in the various passbands for the available subsamples. Since the color information is incomplete, we determined the FP for the same subsamples in Gunn r. All determinations were done with the zero points given by the relative distances

TABLE 5
THE FP IN THE OTHER PASSBANDS

| Passband | $N_E$ | $N_{S0}$ | The FP | rms | $\Delta$(E–S0) |
|---|---|---|---|---|---|
| Gunn g | 45 | 64 | $1.16\log\sigma - 0.76\log<I>_e + \gamma_{cl}$<br>$\pm 0.10\quad\pm 0.04$ | 0.090 | $-0.001 \pm 0.017$ |
| Gunn r selected as Gunn g | 45 | 64 | $1.20\log\sigma - 0.81\log<I>_e + \gamma_{cl}$<br>$\pm 0.09\quad\pm 0.04$ | 0.087 | |
| Johnson B | 46 | 45 | $1.20\log\sigma - 0.83\log<I>_e + \gamma_{cl}$<br>$\pm 0.06\quad\pm 0.02$ | 0.071 | $0.010 \pm 0.015$ |
| Gunn r selected as Johnson B | 46 | 45 | $1.23\log\sigma - 0.83\log<I>_e + \gamma_{cl}$<br>$\pm 0.05\quad\pm 0.02$ | 0.069 | |
| Johnson U | 15 | 26 | $1.08\log\sigma - 0.85\log<I>_e + \gamma_{cl}$<br>$\pm 0.08\quad\pm 0.05$ | 0.067 | $0.001 \pm 0.021$ |
| Gunn r selected as Johnson U | 15 | 26 | $1.25\log\sigma - 0.86\log<I>_e + \gamma_{cl}$<br>$\pm 0.08\quad\pm 0.05$ | 0.067 | |

NOTE.—$\Delta$(E–S0) is the difference between the median zero points for E and S0 galaxies.

derived from Eq. 1. The relations were derived with the same fitting technique as for the FP in Gunn r, e.g. orthogonal fits with the sum of the absolute residuals perpendicular to the fitted relations minimized. Table 5 lists the derived relations together with the rms scatter and the number of galaxies included in the determinations.

The coefficient for $\log<I>_e$ does not depend on the passband. The coefficient for $\log\sigma$ decreases with decreasing wavelength. This implies that the M/L ratio depends stronger on velocity dispersion and mass for the bluer passbands. Our results are in agreement with the results presented by Djorgovski & Santiago (1993). The change of the coefficient for $\log\sigma$ is expected from the observation that the color and $\log\sigma$ are correlated. The rms scatter of the FP does not depend on the passband. This is somewhat surprising, as the influence of starbursts, or age variations, is higher in the bluer passbands. However, if variations in age are linked to variations in the metallicity as $\Delta$[Fe/H]$\approx -2/3\Delta\log t$ at a given velocity dispersion, as suggested by Worthey et al. (1995), then the scatter for the FP is nearly independent of the passband for the passbands we have used. We also note that Pahre, Djorgovski & de Carvalho (1995) found a very similar scatter for the FP derived from photometry in the near infrared K-band and in the optical. Further, we conclude that the intrinsic absorption in E and S0 galaxies cannot be a dominating source of the scatter, since this would give larger scatter for shorter wavelengths.

## 7 THE FP AS A DISTANCE DETERMINATOR

For some choices of coefficients for the FP the residuals correlate strongly with the absolute magnitude. As mentioned in Sect. 3.4 this can bias the distance estimates. Some published values of the coefficients for the FP (e.g. Djorgovski & Davies 1987; Bender et al. 1992; Guzman et al. 1993; Saglia et al. 1993) can produce biases on the order of 5%, if there is a factor two difference between the median luminosity of the galaxies in the observed clusters. For some of the clusters studied by Lynden-Bell et al. (1988) the median luminosity difference of the observed galaxies relative to the galaxies observed in the Coma cluster is a factor 30. Thus, the bias in the distances can be as large as 10-20%. This effect can therefore not be ignored.

For our sample the FP as given in Eq. 1 gives biases smaller than 1%. Thus, we have used this version of the FP to derive relative distances and peculiar velocities for the eleven clusters. The results are given in Table 4. We have normalized the zero point to that of the Coma cluster, for which we assumed a "velocity" distance of 7200km s$^{-1}$. The distances were corrected for cosmological effects following Lynden-Bell et al. (1988). The limiting magnitude of our sample varies from cluster to cluster. We have tested the effect of this by applying a limit of $-20\overset{m}{.}9$ in the total absolute magnitude in Gunn r on all clusters. The differences between the distances derived from this sub-sample and those given in Table 4 are all within $\pm 4\%$, and the mean of the differences is zero. Thus, the variations in the limiting magnitude seem not to have large effects on the derived distances. The distances in Table 4 are not corrected for Malmquist bias. For a uniform density distribution the Malmquist bias for clusters with seven or more observed galaxies is less than 1.5% (Strauss & Willick 1995). The distance to Grm15, which has only four observed galaxies, is subject to a larger bias. Though a uniform density distribution is not realistic this gives an estimate of the size of Malmquist bias to expect. We refer to Strauss & Willick (1995) for an extensive discussion of selection effects and Malmquist bias.

We have further attempted to include a correction for the offset in the Mg$_2$-$\sigma$ relation. As mentioned in Sect. 6.1 the coefficient for $\Delta$Mg$_2$ is most likely smaller than one. Table 4 lists distances and peculiar velocities with the logarithm of the distances offset $<\Delta$Mg$_2$(Coma)$> - <\Delta$Mg$_2>$.

The peculiar velocities we find are generally small. This gives us confidence that the FP is universal to a high degree.

The highest peculiar velocity we find is for S639: 1300 km s$^{-1}$ for the FP. The inclusion of the $<\Delta Mg_2>$ term lowers this peculiar motion to 890 km s$^{-1}$. It has been argued before by Gregg (1992, 1995) and Guzmán et al. (1992) that a $Mg_2$ term should be included to correct for subtle differences in the stellar population. We confirm here that "outlying" peculiar velocities are significantly reduced by inclusion of a $Mg_2$ term. This result can be used in two different ways: one can use the residual from the $Mg_2$-$\sigma$ relation to reject those clusters which show systematic deviations. Alternatively, one can try to correct for the differences by including $\Delta Mg_2$ in the FP. We should note, however, that the optimal coefficient for the correction is not well known.

## 7.1 The $D_n$-$\sigma$ relation

The FP was originally used in the form of the $D_n$-$\sigma$ relation (Dressler et al. 1987b):

$$\log D_n = a \log \sigma + b \qquad (10)$$

We have derived the $D_n$-$\sigma$ relation as the orthogonal fit with the sum of the absolute residuals minimized. The zero points for the clusters were left free as for the determination of the FP in Gunn r. For $D_n$ derived from photometry in Gunn r we find

$$\begin{array}{r}\log D_n = \quad 1.32 \log \sigma + b_{cl} \\ \pm 0.07 \end{array} \qquad (11)$$

The rms scatter in $\log D_n$ is 0.088. For galaxies with velocity dispersion larger than 100 km s$^{-1}$ the scatter is 0.076.

We cannot find a form of the $D_n$-$\sigma$ relation for which the residuals do not correlate with absolute magnitude. We find the same effect for the data set used by Lynden-Bell et al. (1988). For our data we find $\Delta \log D_n = 0.028 \log L - 0.29$. We correct for this correlation by subtracting a term $0.028 \log[L_{median}/L_{median}(Coma)]$ from the median $\Delta \log D_n$ for each cluster. $L_{median}$ is the median absolute luminosity of the galaxies in a clusters.

The resulting peculiar motions are given in Table 4. They are very similar to those derived directly with the FP. The scatter is also similar for the FP and the $D_n$-$\sigma$ relation. This is different from our earlier results on E galaxies in Coma, where the FP produced a much lower scatter (Paper I). The use of velocity dispersions with higher formal errors cannot fully explain why we do not see a large difference in scatter for the FP and the $D_n$-$\sigma$ relation for the large sample studied here. The reason is presumably that other sources than the non-linearity of the $D_n$-$\sigma$ relation dominate the intrinsic scatter, e.g. differences in the stellar populations.

## 8 CONCLUSIONS

We have analyzed the FP for E and S0 galaxies in eleven clusters. We find that the galaxies lie in a plane defined by $\log r_e = 1.24 \log \sigma - 0.82 \log <I>_e + \gamma$, with an rms scatter in $\log r_e$ of 0.084. For galaxies with velocity dispersion larger than 100 km s$^{-1}$ the scatter is 0.073. The FPs derived for the E and the S0 galaxies separately are not significantly different. The clusters investigated in this work include poor groups as Doradus and S639, as well as the rich clusters Coma and DC2345-28. The galaxies occupy a limited area within the FP, which is only partly due to selection effects. There is a possible hint that the FP is slightly curved for highly luminous galaxies, but the effect is not very strong.

The FP implies a strong dependence of the M/L ratio on structural parameters. The scatter of the M/L ratio as a function of $\log r_e$ and $\log \sigma$ is 23%. This systematic variation of M/L ratio with structural parameters is presumable driven by the stellar population. There is a relation between the $Mg_2$ index and the M/L ratio, but with higher scatter than the relation between the M/L ratio and the structural parameters. This might be caused by the same age and/or metallicity variations that cause the scatter in the $Mg_2$-$\sigma$ relation.

The scatter of the FP is twice as high as the expected scatter due to measurement errors. This suggests that there is an unknown source of internal scatter. We have tested whether any of the other observables might be related to this source, but could not find any obvious candidates. As long as this source is not identified, it remains uncertain whether this source might produce systematic errors in distances derived from the FP. There are indications that the intrinsic scatter for certain samples of galaxies is very low, e.g. E galaxies in Coma and galaxies brighter than absolute magnitude $-23.^{\mathrm{m}}1$ in Gunn r. Thus, the distribution perpendicular to the FP may not be the same for all E and S0 galaxies.

The data do not show any significant differences in the coefficients for the FP for the different clusters. There are no significant correlations between the coefficients and cluster properties such as cluster velocity dispersion and temperature of the X-ray gas. The comparison is still difficult, because the galaxies have not been selected to the same absolute magnitude in each cluster. Furthermore, the samples are too small to get good accuracy on the derived coefficients.

We analyzed the correlations between the residuals for the FP and the morphology of the galaxies. Models were constructed of galaxies with disks and bulges, and were used to predict the relations between the residuals and the isophotal shapes and relative disk luminosities. No strong correlations were found in the current data set. This indicates that other sources than disks and projection effects dominate the scatter.

The existence of the FP implies a strong dependence of the stellar population on the structural parameters. The existence of the $Mg_2$-$\sigma$ relation confirms this. The residuals for the $Mg_2$-$\sigma$ relation do not correlate with the residuals for the FP. However, the $Mg_2$-$\sigma$ relation appears to depend on the cluster environment. We find that the residuals for the relation correlate with the projected cluster surface density, as measured by $\sigma^2_{cluster}/R$, where $R$ is the distance of the galaxy from the cluster center. There are indications that the residuals for the FP may also be related to the projected density. However, it is not yet clear whether this effect can cause systematic errors in the distance determinations.

The FP was also derived from photometry in Gunn g, Johnson B, and Johnson U. The scatter does not depend on the passband. This may be the case if variations in the age are linked to variations in the metallicity, as suggested by Worthey et al. (1995).

The use of the FP for distance determinations is explored. We find realistic values for the distances for most clusters if we use the FP in its original form, without any inclusion of $Mg_2$. There is one notable outlier, the cluster S639 for wich we find a peculiar velocity of $1300 \pm 360$ km s$^{-1}$. If we correct for the offset in the $Mg_2$-$\sigma$ relation, $<\Delta Mg_2>$, the peculiar motion of this cluster is reduced to 890km s$^{-1}$. We conclude that the FP in its original form may produce spurious distances for some clusters, but that these clusters can be "flagged" by the residuals from the $Mg_2$-$\sigma$ relation. This confirms earlier results by Gregg (1992, 1995), and Guzmán et al. (1992).

Other indices, such as the H$\beta$ line strength, are needed to explore the causes for the subtle effects with stellar population. Work on this topic is now in progress.

If the offsets in the $Mg_2$-$\sigma$ relation we have found are confirmed with future observations it indicates that not all cluster E and S0 galaxies are drawn from the same probability function, where this probability function is a function of characteristic parameters for the galaxies. External parameters related to the cluster properties may play a role. This also implies that the assumption that all cluster E and S0 galaxies follow the same FP may not be valid, which complicates the use of relations like the FP for determinations of distances and peculiar velocities.

Acknowledgements: The Danish Board for Astronomical Research and the European Southern Observatory are acknowledged for assigning observing time for this project and for financial support. D. Burstein, and R. Guzmán are thanked for giving us data in computer readable form. We thank G. Worthey for supplying his stellar population models. Support for this work was provided by NASA through grant number HF-1016.01.91A to MF and grant number HF-01073.01.94A to IJ, both grants from the Space Telescope Science Institute, which is operated by the Association of Universities for Research in Astronomy, Inc., under NASA contract NAS5-26555.